# Imaging de Haas-van Alphen quantum oscillations and milli-Tesla pseudomagnetic fields

Haibiao Zhou[1#], Nadav Auerbach[1#], Matan Uzan[1#], Yaozhang Zhou[1], Nasrin Banu[1], Weifeng Zhi[1], Martin E. Huber[2], Kenji Watanabe[3], Takashi Taniguchi[4], Yuri Myasoedov[1], Binghai Yan[1] and Eli Zeldov[1*]

[1]Department of Condensed Matter Physics, Weizmann Institute of Science, Rehovot 7610001, Israel

[2]Departments of Physics and Electrical Engineering, University of Colorado Denver, Denver, CO, USA

[3]Research Center for Functional Materials, National Institute for Materials Science, 1-1 Namiki, Tsukuba 305-0044, Japan

[4]International Center for Materials Nanoarchitectonics, National Institute for Materials Science, 1-1 Namiki, Tsukuba 305-0044, Japan

# These authors contributed equally

* eli.zeldov@weizmann.ac.il

A unique attribute of atomically thin quantum materials is the in-situ tunability of their electronic band structure by externally controllable parameters like electrostatic doping, electric field, strain, electron interactions, and displacement or twisting of atomic layers. This unparalleled control of the electronic bands has led to the discovery of a plethora of exotic emergent phenomena (*1*). But despite its key role, there is currently no versatile method for mapping the local band structure in advanced 2D materials devices in which the active layer is commonly embedded in various insulating layers and metallic gates. Utilizing a scanning superconducting quantum interference device, we image the de Haas-van Alphen quantum oscillations in a model system, the Bernal-stacked trilayer graphene with dual gates, which displays multiple highly-tunable bands (*2–4*). By resolving thermodynamic quantum oscillations spanning over 100 Landau levels in low magnetic fields, we reconstruct the band structure and its controllable evolution with the displacement field with unprecedented precision and spatial resolution of 150 nm. Moreover, by developing Landau level interferometry, we reveal shear-strain-induced pseudomagnetic fields and map their spatial dependence. In contrast to artificially-induced large strain, which leads to pseudomagnetic fields of hundreds of Tesla (*5–7*), we detect naturally occurring pseudomagnetic fields as low as 1 mT corresponding to graphene twisting by just 1 millidegree over one µm distance, two orders of magnitude lower than the typical angle disorder in high-quality twisted bilayer graphene devices (*8–11*). This ability to resolve the local band structure and strain on the nanoscale opens the door to the characterization and utilization of tunable band engineering in practical van der Waals devices.

Determining the band structure (BS) and the Fermi surface is a crucial step in understanding and utilizing the electronic properties of materials. The most sensitive canonic method for mapping the BS of bulk metals and semiconductors is measurement of the de Haas-van Alphen (dHvA) oscillations (*12*). In this quantum mechanical effect, in the presence of magnetic field $B$, electrons coherently circulate in closed electronic orbits along the Fermi surface, giving rise to quantum oscillations (QOs) in the grand thermodynamic potential $\Omega$ and in the associated magnetization $M = -\partial\Omega/\partial B$ (*12*). In 2D systems these oscillations are described by the formation of discrete set of Landau energy levels (LLs) with sharp peaks in the density of states (DOS). Charge carriers orbiting in the metallic LL states give rise to diamagnetic response, while ground-state currents flowing in the gapped edge states contribute to paramagnetic magnetization, giving rise to magnetization oscillations with either magnetic field or carrier density (*12*). Since the measured total magnetic moment scales with sample volume, observation of dHvA effect in 2D systems has been challenging (*13*, *14*). Instead, non-thermodynamic Shubnikov–de Haas (SdH) oscillations in the charge transport coefficients is the benchmark characterization tool for atomic layer materials (*15*).

The recent advances in fabrication of multilayer stacked and twisted atomic layer devices based on van der Waals (vdW) materials have provided a fertile ground for a plethora of novel strongly correlated electronic phases which are rare in bulk materials, including tunable correlated insulators (*16*), orbital magnetism (*17*–*19*), integer and fractional Chern insulators (*20*–*23*), unconventional superconductivity (*24*, *25*), and excitonic superfluidity (*26*, *27*). Through material selection, stacking order, and twist angle, a vast variety of vdW structures with different properties can be created. Their BS is strongly dependent on the stacking geometry and can be further manipulated by varying transverse electric field, magnetic field, strain, or pressure. This has ushered in a new era in materials research and offers numerous opportunities for applications in next-generation electronics.

The experimental investigation of the structure of the Fermi surface in micron-sized vdW devices is currently based predominantly on the detection of quantum oscillations using SdH effect (*15*, *28*) and capacitance oscillations (*4*, *25*). However, various types of disorder, such as charge inhomogeneity, twist angle disorder, and strain, widely exist in these samples (*9*, *29*, *30*) resulting in spatial variations in the BS, while the aforementioned methods lack spatial information. While several scanning probe techniques, including scanning tunneling microscopy (*31*, *32*) and single electron transistors (*22*, *33*), are powerful probes of the local electronic properties, the former requires the electron layers to be exposed to vacuum as in photoemission studies, and neither of them can be used for devices encapsulated with a metallic top gate required for application of displacement fields. Development of a universal tool for measurement of the local BS in the diverse family of 2D quantum materials is thus highly desirable.

Strain is a particularly interesting but difficult to characterize tunable parameter, and it should be ubiquitous in vdW devices due to their high mechanical flexibility. In addition to changing the BS and breaking of crystal symmetries, it creates pseudomagnetic fields (PMFs) due to the valley degree of freedom in hexagonal vdW materials (*34*). If sufficiently large, these gauge fields may form effective LLs with sharp peaks in DOS that strongly vary in space and profoundly alter the transport properties of the electronic devices (*5*). PMFs of tens to hundreds of tesla have been observed in artificially strained nanostructures of monolayer graphene (*6*, *7*), bilayer graphene (*35*), twisted homo- (*36*) and hetero-structures (*37*), and folding-formed trilayer graphene (*38*). Yet, the PMFs due to strain naturally forming during the fabrication process have not been observed so far.

Utilizing scanning SQUID-on-tip (SOT) microscopy (*39*), we imaged the dHvA effect in hBN encapsulated dual-gated Bernal-stacked trilayer graphene (TLG) (Fig. 1a). The high magnetic sensitivity of the SOT allows imaging of the QOs at low fields resolving the multi-band electronic structure with sub-meV energy resolution and unperturbed by elevated magnetic fields. The quantitative information provided by the thermodynamic oscillations allows unique high-precision derivation of the tight-binding hopping parameters and accurate reconstruction of the tunable band hybridization induced by the displacement field. Moreover, the nanoscale spatial resolution allows detailed quantitative study of the spatial variations of the QOs over the entire device, revealing the presence of PMFs of mT magnitude in micron-sized domains.

**Band structure of ABA graphene**

The Bernal-stacked ABA TLG is the minimal graphene structure that requires the full set of parameters in the Slonczewski-Weiss-McClure (SWMc) tight-binding model (*40*) proposed for describing the BS of graphite (Fig. 1e) with six hopping parameters $\gamma_i$ ($i = 0$ to $5$), on-site energy difference $\delta$ between the A and B sublattices, potential difference $\Delta_1$ between the top and bottom graphene layers induced by the applied displacement field $D$, and $\Delta_2$ potential that describes the non-uniform charge distribution between the middle and outer layers. The influence of these individual parameters on the BS is shown in Extended Data Fig. 5.

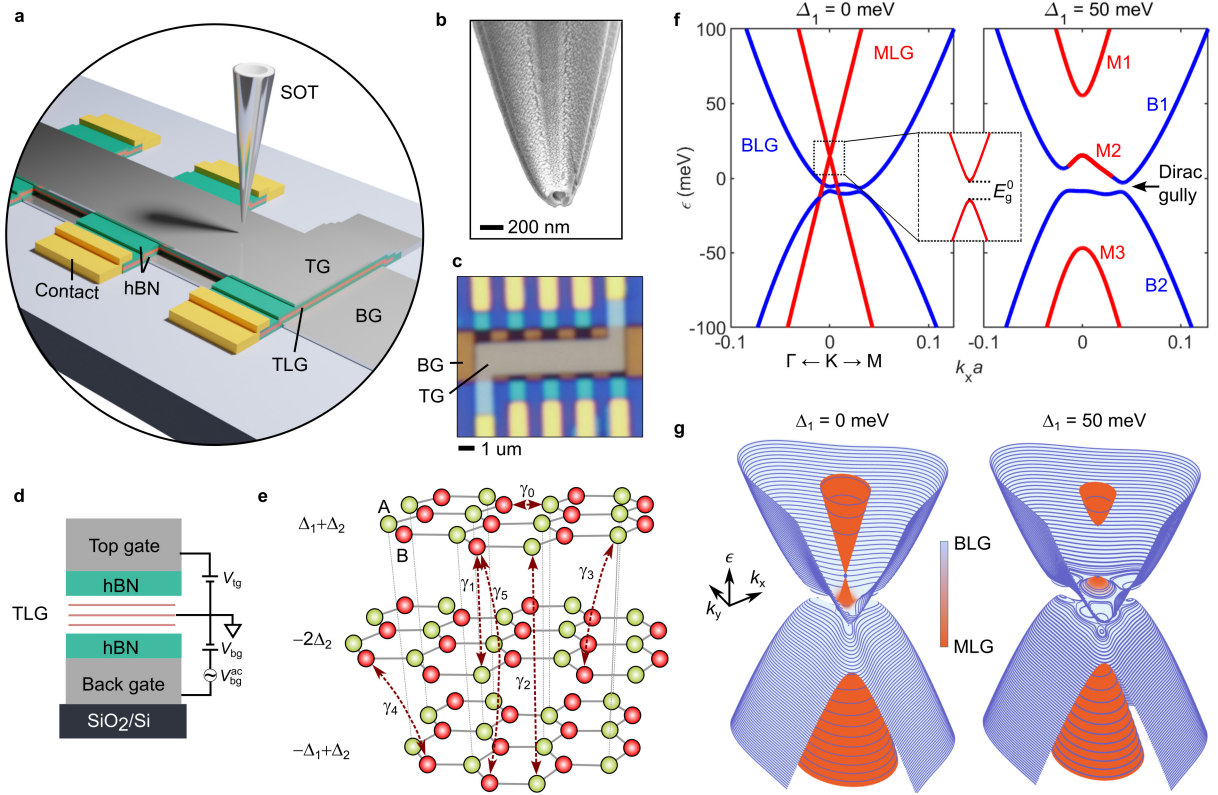

**Fig. 1. Experimental setup and ABA graphene band structure**. **a,** Schematic sample structure with trilayer graphene (TLG) encapsulated by hBN and top (TG) and back (BG) Pt gates, with scanning SQUID-on-tip (SOT). **b,** Scanning electron microscope image of the Indium SOT. **c,** Optical image of the TLG device. **d,** The stacking geometry of the device with indicated voltages applied to the gates. **e,** Atomic structure of ABA graphene with indicated SWMc parameters. **f,** The BS of ABA TLG for zero displacement field ($\Delta_1 = 0$ meV) and for $\Delta_1 = 50$ meV. Inset: a small Dirac gap $E_g^0$ is present in the MLG band at $\Delta_1 = 0$ meV which grows rapidly with $\Delta_1$. **g,** 3D rendering of the BS reconstructed from the dHvA oscillations with overlaid contours of the calculated LLs. The LLs are shown for $B_a = 1$ T for clarity. At our $B_a = 320$ mT the LLs are three times denser. The color map represents the wavefunction projection onto the MLG (red) and BLG-like (blue) bands.

Due to the mirror symmetry of the crystal structure, the bands can be decomposed into a monolayer graphene-like (MLG) Dirac band and a bilayer graphene-like (BLG) quadratic band, with a relative energy shift between them (Fig. 1f left). A transverse displacement field breaks the mirror symmetry and induces strong hybridization between the two bands, causing multiple Lifshitz transitions. At high $D$, the Dirac band is divided into three sections, which we label M1, M2 and M3 (Fig. 1f right). The M1 and M3 bands remain separated from the BLG-like bands B1 and B2, while M2 merges with B1 and evolves into gapped mini-Dirac cones (Dirac gullies) at the bottom of B1. The gullies have a three-fold rotational symmetry leading to various possible quantum Hall ferromagnetic and nematic states (*41*, *42*).

Previous studies have explored SdH and capacitance oscillations in TLG at elevated fields (*3*, *43*, *44*) to determine the BS and search for broken-symmetry states (*4*, *45*, *46*), resulting in a wide span of derived SWMc parameters (Table 1 in Methods). However, the details of the ABA graphene BS are still under debate. In particular, the size and the sign of the small gap in the Dirac band at $D = 0$ (Fig. 1f inset), $E_g^0 = \delta + \frac{\gamma_2 - \gamma_5}{2}$, is controversial (*46*). In addition, it was predicted that the trigonal warping due to non-vanishing $\gamma_3$ breaks the rotational symmetry of the BS with significant consequences on the LL structure, resulting in level anticrossings, which occur between a given MLG LL and every third BLG LL (*42*, *47*). Though some single anticrossings were reported (*46*), the predicted periodicity has not been observed directly. Moreover, some gully-polarized states have been reported in high magnetic fields (*4*), but gully coherence at low fields remains an open question.

To address these questions, a technique capable of probing the BS with high energy resolution and at low magnetic fields is required. The unique advantage of the dHvA effect is that the QOs are fully described by thermodynamic quantities which can be readily derived directly from BS calculations, thus allowing detailed quantitative analysis, which is not possible in other methods like SdH oscillations.

**Nanoscale magnetic imaging and results**

The TLG device (Fig. 1c) was fabricated using the common dry transfer method, with the graphene flake sandwiched between two Pt gates, separated by $\cong$30 nm of hBN (Methods), as shown schematically in Fig. 1a. Transport measurements of $R_{xx}$ vs. the carrier density $n$ and the applied magnetic field $B_a$ show a Landau fan with several LL crossings (Extended Data Fig. 1), similar to previous reports (*46*). The local measurements of the dHvA oscillations were performed in a home-built cryogen-free scanning SOT system at $T \cong 160$ mK (Methods). The Indium SOT (Fig. 1b) had an effective diameter of 150 nm and a magnetic sensitivity of 20 nT/Hz$^{1/2}$ (Methods). Voltages $V_{tg}$ and $V_{bg}$ applied to the top and bottom gates (Fig. 1d) are used to control the carrier density $n$ and the displacement field $D$. A small *ac* voltage $V_{bg}^{ac}$ at a frequency of about 1.8 kHz modulates $n$ by $n_{ac}$, and the corresponding induced local *ac* magnetic field $B_z^{ac}$ is measured by the SOT as a function of $n$, $D$, and the position across the sample. $B_z^{ac}$ reflects the differential change $m_z$ in the local orbital magnetization $M_z$, $m_z = \partial M_z / \partial n$.

We first describe the QOs at a single point by positioning the SOT statically at a height of $h \cong 150$ nm above the graphene. Figure 2c shows the dHvA oscillations acquired at $B_a = 320$ mT as a function of $n$ and $D$, focusing on low carrier densities between $-1.2 \times 10^{12}$ and $2.3 \times 10^{12}$ cm$^{-2}$. A line cut of the data at $D = 0$ V/nm is shown in Fig. 2a. Remarkably, in this relatively small range of $n$ we observe over 100 LLs, in sharp contrast to transport measurements (Extended Data Fig. 1) where almost no SdH oscillations can be discerned at such low $B_a$. Moreover, we can resolve dHvA oscillations at fields as low as 40 mT as shown in Extended Data Fig. 3. To the best of our knowledge, this is the lowest magnetic field at which QOs have been observed in 2D systems. As illustrated in Fig. 1g, mapping the very dense LLs provides a unique quantitative tool for reconstruction of the BS with unprecedented precision, allowing us to derive high accuracy SWMc parameters as described in Methods and summarized in Table 1.

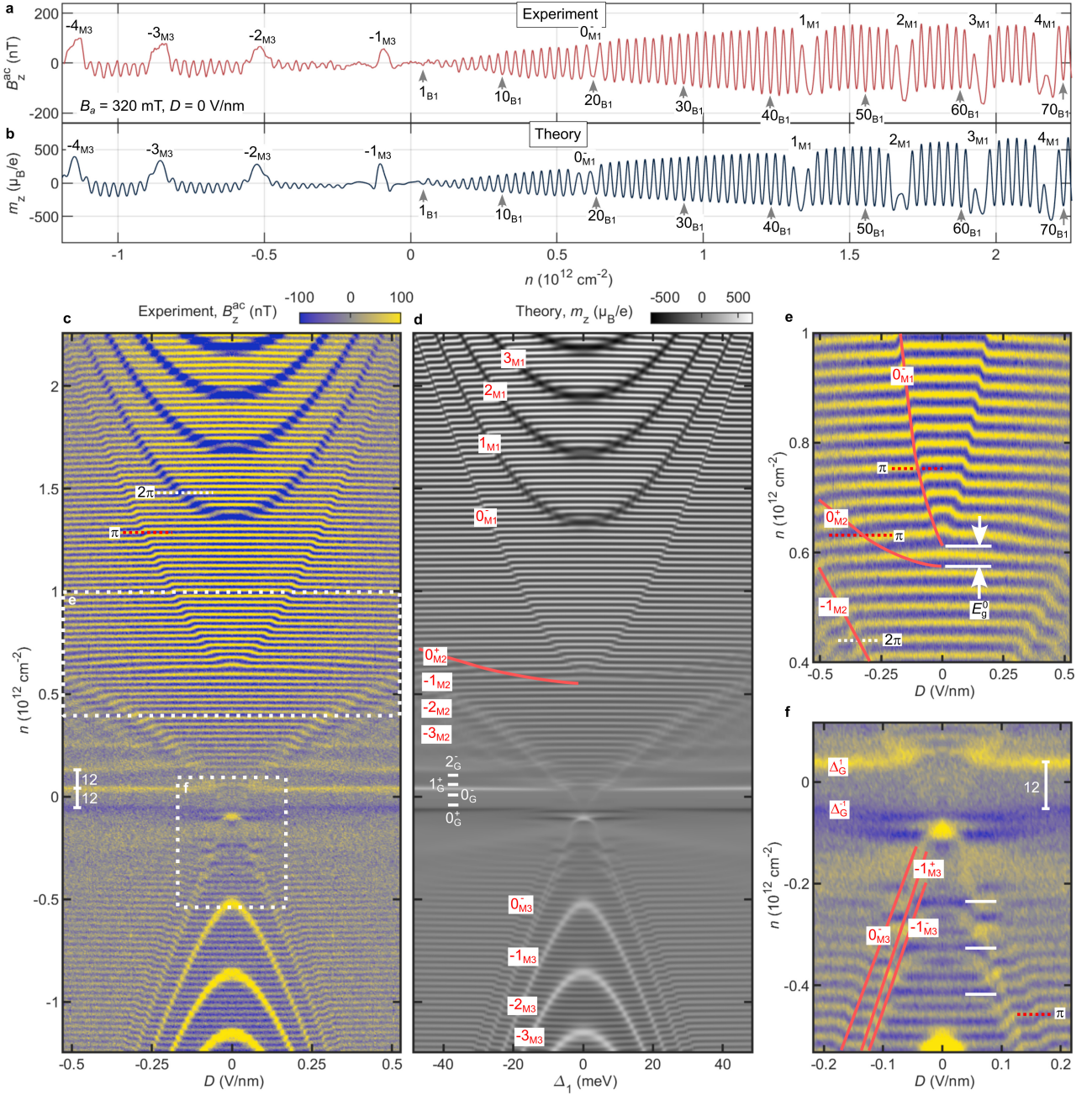

**Fig. 2. Measurement of the dHvA effect in ABA graphene.** **a**, The measured local magnetic QOs signal $B_z^{ac}$ as a function of $n$ at 160 mK at a fixed location in the interior of the sample at $B_a = 320$ mT and $D = 0$ V/nm using $V_{bg}^{ac} = 8$ mV rms. Some indices of the LLs in M1, B1, and M3 bands are indicated. **b**, The calculated dHvA differential magnetization $m_z$ at $D = 0$ V/nm using the derived BS. **c**, The measured $B_z^{ac}$ vs. $n$ and $D$. Crossings between the four-fold degenerate BLG LLs (horizontal yellow/blue lines) and the 0$^{th}$ $K^-$ valley MLG LL, $0_{M1}^-$ results in a $\pi$ shift (red dotted line), while crossing with the higher MLG LLs introduces a $2\pi$ shift (white dotted line). The white vertical bars indicate the 12-fold degeneracy of the LLs in the gullies. **d**, Calculated $m_z(n, D)$ using the fitted SWMc tight-binding parameters with Dingle broadening of 0.3 meV. The MLG LLs in M1, M2, and M3 bands and the gully LLs are labeled. **e**, Zoom-in of the measured $B_z^{ac}(n, D)$ in the vicinity of MLG Dirac gap $E_g$ with marked MLG LLs (red). The $\pi$ shifts at crossings between BLG and $0_{M2}^+$ and $0_{M1}^-$ valley-polarized MLG LLs and $2\pi$ shift at crossing of valley degenerate $-1_{M2}$ LL are indicated. **f**, Zoom-in of $B_z^{ac}(n, D)$ near the top of the BLG valence band B2 showing anticrossings between the MLG and BLG LLs. The white horizontal bars indicate enlarged anticrossing gaps for every third BLG LL (see Extended Data Fig. 8).

Figures 2b,d present the dHvA oscillations calculated from the BS derived using the fitted SWMc parameters (Methods), showing remarkable qualitative and quantitative agreement with the experimental data. Such accurate comparison between the theory and experiment is possible because the thermodynamic QOs in the magnetization can be calculated quantitatively from the BS. The observed QOs can be classified by five prominent sets of LLs as follows:

**B1 and B2 bands**. As the DOS in the BLG bands is much larger than in MLG bands, the LLs in B1 and B2 bands appear as dense horizontal lines in Figs. 2c,d. At low fields the LLs are four-fold valley and spin degenerate and disperse as $\sim \pm\sqrt{n_B(n_B-1)}B_a$ where $n_B$ is the LL index in the BLG bands (positive for electrons and negative for holes). At our $B_a = 320$ mT, the energy spacing between the BLG LLs is about 1 meV in B1 band and 0.6 meV in B2, defining our energy resolution of better than 0.6 meV.

**M1 band**. Due to the low DOS of the MLG bands and approximate linear dispersion, the MLG LLs are much sparser and disperse as $\sim\pm\sqrt{|n_M|B_a}$. Displacement field opens a large gap $E_g$ between the MLG sections (Figs. 1f,g) resulting in parabolic-like upturn of M1 LLs with $D$ in Figs. 2c,d. When the B1 LLs are crossed by an M1 LL, there is a shift in the position of B1 LLs vs. $n$ because the M1 LL has to be filled before the following B1 LLs can be occupied. The size of the shift depends on the degeneracy of the M1 LL. All the higher M1 LLs are four-fold degenerate, like the B1 LLs. As a result, at the crossing points, the B1 LLs show a phase shift of $2\pi$ as marked by the dotted white line in Figs. 1c,e. The 0th MLG LLs are fundamentally different. Due to the topological nature of the Dirac point and the associated Berry phase, the 0th LL are valley polarized with the $K^-$ 0th LL, $0^-_{M1}$, residing at the bottom of M1 band while the $K^+$ 0th LL, $0^+_{M2}$, is pinned to the top of M2 band. Since the 0th LLs are two-fold spin degenerate, their crossing with the four-fold degenerate B1 LLs gives rise to a phase shift of $\pi$ rather than $2\pi$, as shown by the red dotted lines in Figs. 1c,e (see also Methods and Extended Data Figs. 6d,e). Moreover, higher M1 LLs display a pronounced negative (dark) signal due to the diamagnetic response in the compressible states (*48*). Remarkably, as opposed to all other LLs, the $0^+_{M1}$ and $0^-_{M2}$ LLs in Figs. 2c-e are invisible showing no diamagnetic signal, and their presence is discerned only as a phase shift in the B1 LLs. This is the result of the Dirac Berry curvature giving rise to 0th LLs pinned to the band extrema with no kinetic energy and no magnetic field dependence, and hence no diamagnetism in their compressible states. In contrast, the incompressible states in the gaps of MLG LLs do show a paramagnetic response (*48*) determined by the gap Chern number $C$ as shown in Extended Data Fig. 3c.

**M2 band**. The hybridization of M2 and B1 bands results in a small M2 hole pocket with total DOS that is too low to accommodate even a single LL at elevated magnetic fields. As a result, the M2 LLs could not be identified in previous studies (*4, 43, 46*). Our low $B_a$ and high sensitivity allow clear resolution of M2 LLs as shown in Figs. 2c-e and Extended Data Fig. 3. In contrast to M1 and M3 LLs that coexist with B1 and B2 bands with the same carrier type, the M2 hole LLs coexist with B1 electron LLs, reflecting a Mexican hat potential. Figure 2e shows a gap $E_g^0$ between the $0^-_{M1}$ and $0^+_{M2}$ LLs at $D=0$, that is comparable to the gap between two B1 LLs ($\cong$ 1 meV). Upon increasing $D$, the $0^-_{M1}$ and $0^+_{M2}$ LLs spread apart rapidly without crossing, clearly indicating that $E_g$ grows continuously without intermediate gap closing, in contrast to previous suggestions (*4, 46*).

**M3 band**. At elevated hole doping the M3 LLs are four-fold degenerate, showing behavior similar to the M1 LLs. At low doping, in contrast, the strong hybridization between M3 and B2 bands gives rise to an unusual valley polarization. This is demonstrated in Fig. 2f by the four-fold degenerate $-1_{M3}$ LL, which splits into valley polarized, spin-degenerate $-1^+_{M3}$ and $-1^-_{M3}$ LLs at low hole doping. This valley splitting process is accompanied by multiple M3 and B2 LLs crossings and anticrossings as described in Extended Data Fig. 8. In particular, BS calculations have predicted anticrossings between BLG LLs and low-index M3 LLs arising from the trigonal warping term $\gamma_3$ (*42*, *47*), giving rise to avoided crossings with enhanced gaps for every third BLG LL. This tripled period of anticrossings with BLG LLs, which has remained unidentified so far, is clearly resolved in our data marked by the white bars in Fig. 2f in excellent agreement with the calculated results in Extended Data Fig. 8. In addition, we resolve also the $0^-_{M3}$ LL which has no diamagnetism and gives rise to a $\pi$ shift in the B2 LLs as shown by the red dotted line in Fig. 2f.

**LLs in the gullies**. Upon increasing $D$, the enhanced hybridization between the MLG and BLG bands and the trigonal warping give rise to formation of three-fold rotationally symmetric Dirac gullies (*42*, *47*) and to highly intriguing evolution of the LLs. The low energy BLG LLs, which are mostly valley degenerate at $D = 0$, undergo valley polarization and regrouping with $D$, forming valley-polarized six-fold degenerate LLs in the gully pockets as shown in Extended Data Fig. 4f. The 0$^{th}$ electron and hole gully LLs, $0^-_G$ and $0^+_G$, display no diamagnetism due to the Berry phase, and the gap between them, $\Delta^0_G$ has zero Chern number $C = 0$. As a result, a zero magnetism strip of width corresponding to 12-fold degeneracy (incompressible gap and $0^-_G$ and $0^+_G$ compressible states) is observed around the charge neutrality point (CNP) in Figs. 2c,d,f at elevated $D$. The positive and negative (yellow and blue) signals above and below the zero-magnetization strip are the paramagnetic responses in the LL gaps $\Delta^1_G$ and $\Delta^{-1}_G$, as marked in Fig. 2f and Extended Data Figs. 4e,f.

The intricate structure of the QOs is highly sensitive to the shape of the BS and its evolution with $D$. Our fine energy resolution thus allows high precision determination of the SWMc parameters and of the Dingle broadening of the LLs as described in Methods.

**LL interferometry and pseudomagnetic fields**

Next, we analyze the QOs over the full range of accessible carrier densities $|n| \lesssim$ 9×10$^{12}$ cm$^{-2}$, which allows resolution of much finer details of the BS and its spatial dependence. Since in this range of $n$ at $B_a =$ 320 mT there are about 500 BLG LLs in addition to over 100 MLG LLs, we focus on the sparser MLG LLs by applying a larger $V^{ac}_{bg}$ excitation, which averages out the BLG QOs and enhances the visibility of the MLG LLs (Methods). Figures 3a-d show the spatial dependence of $B^{ac}_z(x, y)$ at several carrier densities $n$, while Fig. 3f presents $B^{ac}_z(x)$ vs. $n$ along the dotted line in Fig. 3a revealing QOs due to the MLG LLs. For $|n| \lesssim$ 3×10$^{12}$ cm$^{-2}$, discussed in Fig. 2, the QOs display relatively high spatial uniformity as shown in Fig. 3d and at the bottom of Fig. 3f, demonstrating high sample quality. At higher carrier densities, however, distinctly different behavior is observed depending on the location as demonstrated in Figs. 3g,h showing the QOs at sites A and B indicated in Fig. 3b. Large parts of the sample, exemplified by site A, show continuous evolution of the amplitude of the QOs with $n$ (Fig. 3g), which matches very well the calculated magnetization oscillations considering the finite Dingle broadening and the finite $V^{ac}_{bg}$ (Methods). In other parts of the sample, however, striking low-frequency beating of the MLG LLs is found as shown in Fig. 3h at site B. Upon decreasing the field to 170 mT, more beating nodes are observed and their position is shifted to lower LL indices (Fig. 3i).

The low-frequency beating of MLG QOs indicates interference of two close oscillation frequencies, usually originating from two bands with very similar dispersion relations. Since the MLG and BLG bands have very different dispersions, the beating cannot arise due to interference between these two bands. It must therefore originate from small symmetry breaking between the four flavors of the MLG band. In Methods we consider various possible mechanisms of such degeneracy lifting including staggered substrate potential, Kekulé distortions, band shifting, Zeeman effects, and spin-orbit coupling, as well as non-symmetry-breaking disorder, and show that they are inconsistent with the observed behavior. In the following, we demonstrate that the interference of the QOs is well described by the natural presence of strain-induced pseudomagnetic fields $B_S$.

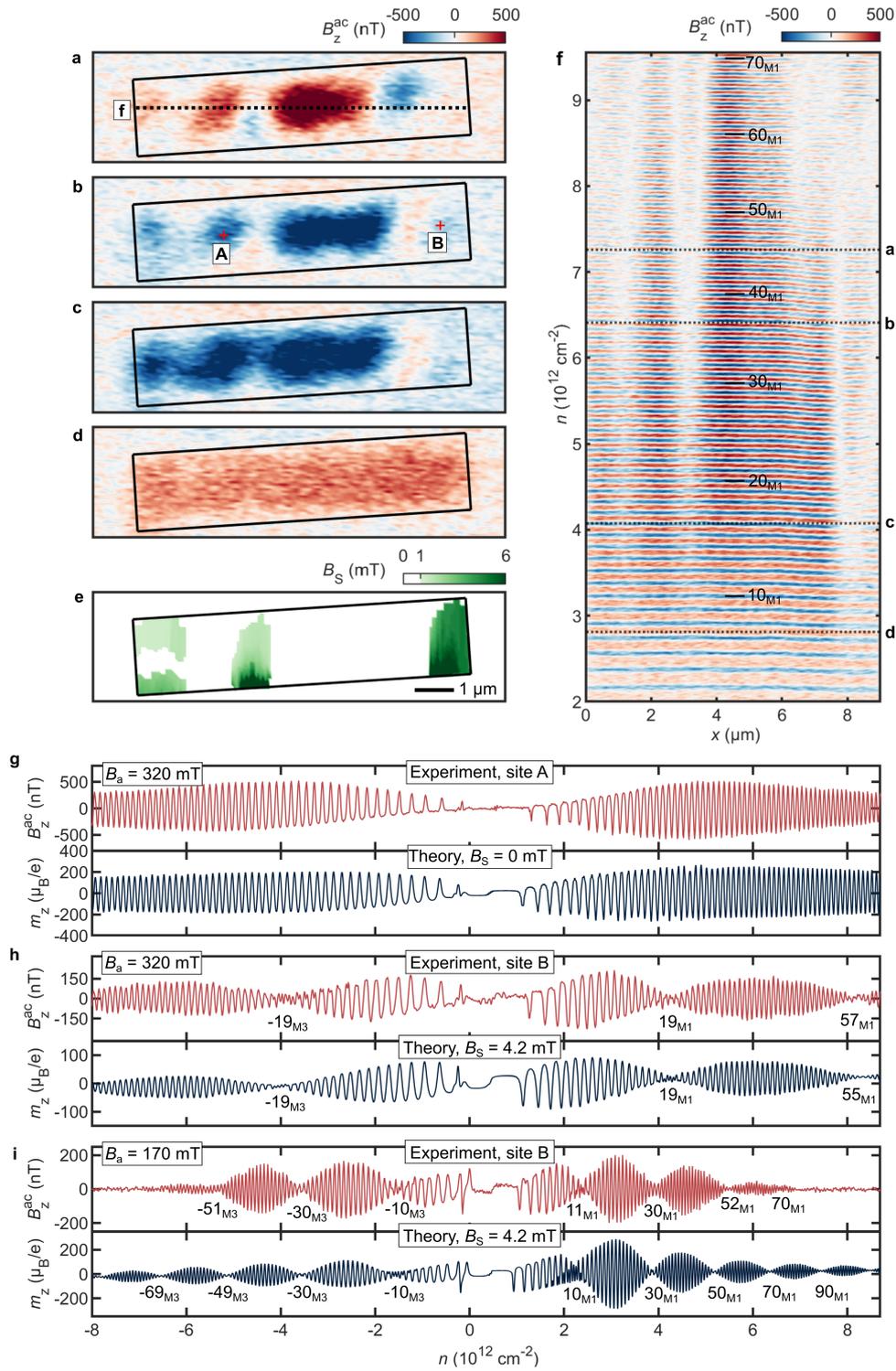

**Fig. 3. Imaging the beating of QOs and mapping the pseudomagnetic fields. a-d,** Spatial imaging of $B_z^{ac}(x,y)$ at $B_a = 320$ mT and $n = 7.26\times10^{12}$ (a), $6.41\times10^{12}$ (b), $4.08\times10^{12}$ (c) and $2.81\times10^{12}$ (d) cm$^{-2}$ corresponding to dotted lines in (f). The black rectangle indicates the boundaries of the TLG. **e,** Map of the derived pseudomagnetic field $B_s$ across the sample. Regions with $B_s$ below our resolution of 1 mT are shaded in white. **f,** Line scans of $B_z^{ac}(x)$ vs. $n$ measured along the dotted line marked in (a) showing QOs due to MLG LLs in M1 band. **g,** The measured QOs due to MLG LLs (upper panel) at location A indicated in (b) and the calculated QOs (bottom panel). **h,** The measured QOs (upper panel) at location B and the calculated QOs (bottom panel) with $B_s = 4.2$ mT. The LL indices at the beating nodes are indicated. **i,** Same as (h) at $B_a = 170$ mT. The applied larger $V_{bg}^{ac} = 10$ mV rms in (a-d) and (i), and 20 mV rms in (f-h) averages out the QOs due to BLG LLs intensifying the visibility of MLG LLs.

Long-wavelength mechanical strain induces an effective gauge field in graphene, which has an opposite sign for the two valleys. Isotropic or uniaxial strains result in zero PMF, whereas non-uniform shear strain gives rise to finite $B_S$ (34). In the presence of magnetic field, the carriers in the $K^+$ and $K^-$ valleys experience effective fields $B_{eff}$ of $B_a + B_S$ and $B_a - B_S$ which causes a relative shift in the LLs leading to interference as illustrated in Fig. 4a. In the MLG Dirac band the LL energies are given by $E_N = \sqrt{2e\hbar v_F^2 B_{eff} N}$, and hence for $B_S \ll B_a$ the energy shift between the LLs in the two valleys is $\delta E_N = \sqrt{2e\hbar v_F^2 N}(\sqrt{B_a + B_S} - \sqrt{B_a - B_S}) \cong B_S\sqrt{2e\hbar v_F^2 N/B_a}$. Since $\delta E_N$ scales with $\sqrt{N}$, the lowest LLs remain essentially fully degenerate. For higher LLs, the relative shift between the valley-polarized LLs grows continuously with energy, giving rise to beating. The first destructive interference occurs when $\delta E_N = \Delta E_N/2$, where $\Delta E_N = E_{N+1} - E_N = \sqrt{2e\hbar v_F^2 B_a}(\sqrt{N+1} - \sqrt{N}) \cong \sqrt{e\hbar v_F^2 B_a/(2N)}$ is the LL energy spacing, resulting in the first beating node at $N = N_b^1 = B_a/(4B_S)$.

Figure 4b shows the theoretical $N_b^1$ dependence on $B_S$ for $B_a = 320$ mT, and the calculated beating patterns of the QOs vs. $B_S$ are presented in Fig. 4d with overlaid traces (red) of the beating nodes $N_b^1$ to $N_b^3$. At the nodes, the $K^+$ and $K^-$ LLs are out of phase, giving rise to amplitude suppression and to a barely visible frequency doubling. In Fig. 3h (red curve) we find that the first node occurs at $N_b^1 = 19$, from which we derive $B_S = B_a/(4N_b^1) = 4.2$ mT. The theoretically calculated QOs in the presence of $B_S = 4.2$ mT (black curve) show excellent agreement with the data and in addition well reproduces the secondary nodes at $-19$ and $57$. Moreover, the experimental evolution of the interference of the QOs with $D$ (Fig. 4e) is well reproduced by the simulations (Fig. 4f) both in the position of all the three beating nodes (red arrows) and in their evolution with $D$ using a single fitting parameter $B_S$. Since the BS changes profoundly with $D$ while the strain-induced PMF should be independent of the BS, the fact that the observed beating is well described by a $D$-independent $B_S$ provides an additional strong support of the model. Furthermore, the PMF should affect all the bands. An important self-consistency check is therefore observation of beating also in the BLG LLs. Extended Data Fig. 9 shows a closer examination of the BLG QOs acquired at the same location B. Beating is clearly observed and well reproduced by simulations using the same set of parameters as in Figs. 3 and 4. Finally, a crucial test of the PMF origin of the interference is the predicted linear dependence of $N_b^1$ on $B_a$, which sets it apart from other possible mechanisms (Methods). Figure 3i (red curve) presents the LL interference measured at a lower field of 170 mT showing multiple beating nodes. The calculated QOs (black curve) show an excellent agreement with the data with the same $B_S$ as derived at 320 mT, confirming the linear $N_b^1$ dependence on $B_a$ as shown in Fig. 4c.

By analyzing the LL interference over the entire sample, we derive a map of $B_S$ as shown in Fig. 3e. In large parts of the sample, $B_S$ is below our resolution of about 1 mT determined by the largest accessible carrier density. We find three regions with characteristic length $L \cong 1$ μm with smoothly varying $B_S$ reaching up to 6 mT. Two types of lattice distortions—finely tuned triaxial strain or arc-like in-plane bending—have been shown theoretically to produce relatively homogenous $B_S$ (5, 49). In our sample it is very likely that the source of the strain is a small arc-like in-plane twist introduced during the fabrication processes (Fig. 4b inset).

An arc segment of length $L$ with a small twist angle $\theta$ in a graphene strip generates $B_S = c\beta\frac{\phi_0}{aL}\theta$ (49), where $a = 0.14$ nm is the graphene interatomic distance, $c \cong 1$ is a numerical constant, $\beta \cong 2$ describes the hopping parameter dependence on $a$, $\phi_0 = h/e$ is the flux quantum, $h$ is the Planck constant, and $e$ is the elementary charge. The measured $1 < B_S < 6$ mT thus corresponds to twisting by $1 < \theta < 6$ millidegree, or equivalently bending radius of $1 < R < 6$ cm and corresponding strain of $8 \cdot 10^{-6} < \bar{u} < 5 \times 10^{-5}$, where $R = L/\theta$ and $\bar{u} = \theta/2$ (*49*). Such minute bending angles and strains should be abundant in exfoliated atomic layer devices. In fact, orders of magnitude larger strains and angle disorder have been reported to occur naturally in twisted and stacked graphene structures (*8–11, 32, 50*).

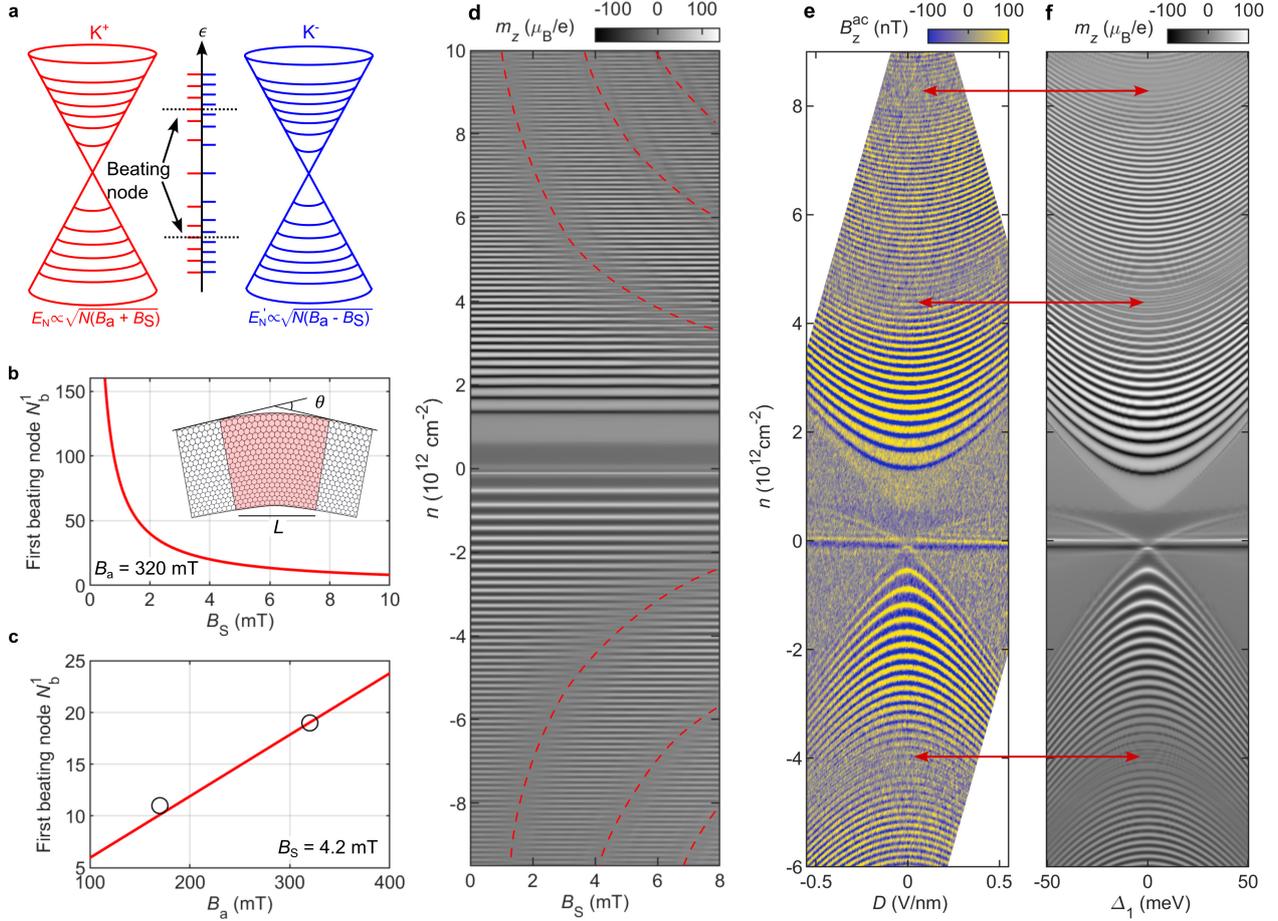

**Fig. 4. LL interferometry of strain-induced pseudomagnetic field**. **a**, Schematic of LL beating in the presence of PMF ($B_S$) in graphene. Only the MLG Dirac bands with LLs are shown for clarity. **b**, The LL index of the first beating node $N_b^1$ in the Dirac band vs. $B_S$ at $B_a = 320$ mT. Inset: schematic of a graphene strip with arc-like bent section (red) of length $L$ and bend angle $\theta$ generating strain-induced $B_S$. The illustrated $\theta$ is greatly exaggerated compared to the maximal derived $\theta \cong 6 \times 10^{-3}$ degree. **c**, Calculated dependence of $N_b^1$ on $B_a$ for $B_S = 4.2$ mT (red). The open circles show the measured $N_b^1$ at $B_a = 170$ and 320 mT. **d**, Calculated QOs in the MLG band vs. $n$ and $B_S$ at $B_a = 320$ mT and $D = 0$ V/nm. The locations of the beating nodes are highlighted in red. **e**, The measured QOs in the MLG bands vs. $n$ and $D$ using $V_{bg}^{ac} = 20$ mV rms showing beating nodes (red arrows). **f**, Calculated QOs vs. $n$ and $D$ at $B_a = 320$ mT and $B_S = 4.2$ mT.

**Discussion**

Our measurements of dHvA effect in low fields have revealed multiple fine features of the BS in a trilayer graphene device, which allow high-precision derivation of the hopping parameters in the SWMc model, while the nanoscale spatial resolution enables mapping the local BS over the entire device. The revealed PMFs in a honeycomb lattice, cause beating of the LLs, which is usually ascribed to sub-band splitting or spin-orbit coupling. The presence of PMFs has important implications for our comprehension of various types of disorder and their impact on strongly correlated states in a wide range of vdW structures. In particular, magic angle twisted bilayer and multilayer graphene are known to be prone to twist angle disorder which strongly affects their properties. Spatial variations in twist angle and strain are understood to cause fluctuations in the bandwidth of the flat bands, in electron interactions, and in formation of symmetry broken states (*51*). Yet, the effects of the accompanying spatially varying PMF have not been investigated systematically. Our findings imply that the typical reported twist angle disorder of 0.1° (*8–11*) will generate $B_S \cong 0.1$ T, which should greatly affect the magnetotransport behavior. Indeed, resolving LLs in transport at low fields has proven to be challenging in twisted devices, which could be attributed to such highly spatially varying PMFs. Moreover, in the presence of magnetic field, the PMF breaks the valley symmetry resulting in different DOS in the two valleys. $B_S$ fluctuations may thus affect the local Stoner instabilities and symmetry breaking mechanisms that lead to the quantum anomalous Hall effect, Chern insulators, and inhomogeneities in the spontaneous orbital magnetization (*18–20*). Finally, the recent development of controllable in-plane bending of graphene ribbons (*52*) opens the door to microscale engineering and exploitation of PMFs. The derived method of high precision determination of the local band structure and imaging of the pseudomagnetic fields provides a powerful tool for characterization and optimization of tunable electronic bands and calls for further investigation of the role of strain-induced gauge fields in the formation of symmetry broken strongly correlated states of matter.

## Methods

### Device fabrication

The hBN-encapsulated ABA graphene heterostructure was fabricated using the dry-transfer method. The graphene flakes were first exfoliated onto a Si/SiO$_2$ (285 nm) substrate. The number of layers in the graphene flakes was determined using Raman microscopy (*53*). Then, the hBN (about 30 nm thick) and the graphene flakes were picked up using a polycarbonate (PC) on a polydimethylsiloxane (PDMS) dome stamp. The stacks were then released onto a pre-annealed Ti (2 nm)/Pt (10 nm) bottom gate, patterned on the Si/SiO$_2$ wafer. The finalized stacks were annealed in vacuum at 500 °C for strain release (*54*). A Ti (2nm)/ Pt (10nm) top gate was then deposited on top of the stack. The one-dimensional contacts were formed by SF$_6$ and O$_2$ plasma etching followed by evaporating Cr (4 nm)/Au (70 nm). Then, the device was etched into a Hall bar geometry. Finally, the device was re-annealed at 350 °C in vacuum. The capacitances per unit area of the back and top gates are $C_{bg}$ = 0.649 x 10$^{12}$ e·cm$^{-2}$·V$^{-1}$, $C_{tg}$ = 0.668 x 10$^{12}$ e·cm$^{-2}$·V$^{-1}$. The top and back gates are used to control the carrier density $n = (C_{bg}V_{bg} + C_{tg}V_{tg})/e$ and the effective transverse displacement field $D = (C_{tg}V_{tg} - C_{bg}V_{bg})/2\varepsilon_0$, where $\varepsilon_0$ is vacuum permittivity. From fitting the experimental QOs to simulations we find that $D = 1$ V/nm corresponds to energy difference between the top and bottom graphene layers of $\Delta_1 = 92$ meV.

### Transport measurements

Transport characterization of ABA graphene devices was carried out using standard lock-in techniques. The $R_{xx}$ shows a peak along the diagonal charge neutrality line which increases with $D$, suggesting gap opening (Extended Data Fig. 1a). The Landau fan shows LL crossings (Extended Data Fig. 1b), consistent with previous reports (*3*, *43*, *44*, *46*, *55–60*). The QOs from MLG band LLs are visible at low fields but the BLG LLs can only be resolved above 0.75 T on the electron side and at significantly higher fields on the hole doping side (Extended Data Fig. 1c).

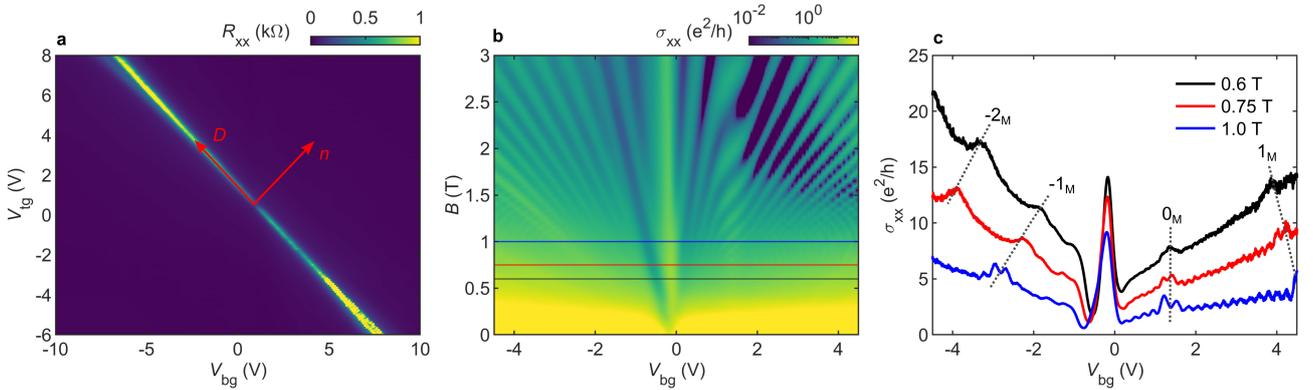

**Extended Data Fig. 1 | Transport characterization of ABA graphene. a,** A dual-gate sweep measurement of $R_{xx}$ at $T = 200$ mK in $B_a = 0$ T in device A described in the main text. **b,** The Landau fan of $\sigma_{xx}$ in device B at $T = 1.67$ K. **c,** The Shubnikov–de Haas oscillations in $\sigma_{xx}$ at 0.6, 0.75 and 1.0 T along the lines marked in (b) with indicated MLG LLs. The BLG LLs are visible only at $B_a \gtrsim 0.75$ T for electron doping.

### SQUID-on-tip measurements and magnetization reconstruction

The local magnetic measurements were conducted in a home-built scanning SQUID-on-tip microscope in a cryogen-free dilution refrigerator (Leiden CF1200) at temperature of 160 to 350 mK (*61*). Indium SOT with a diameter of about 150 nm was fabricated as described previously (*39*, *62*, *63*). The SOT readout circuit is based on SQUID series array amplifier (*64*, *65*). The SOT is attached to a quartz tuning fork vibrating at about 32.8 kHz (Model TB38, HMI Frequency Technology), which is used as a force sensor for tip height control (66). The scanning height was about 150 nm above the ABA graphene. An ac voltage $V_{bg}^{ac}$ at frequency of about 1.8 kHz was applied to the backgate to modulate the carrier density by $n_{ac} = C_{bg}V_{bg}^{ac}/e$. A lock-in amplifier was used to measure the corresponding local $B_z^{ac}$ by the scanning SOT. The $B_z^{ac}$ data were symmetrized with respect to the displacement field $D$ where applicable.

The 2D $B_z^{ac}(x,y)$ images were used to reconstruct the magnetization $m_z(x,y)$ using numerical inversion procedure described in Ref. (67) as shown in Extended Data Fig. 2. Since the reconstruction of $m_z$ requires 2D $B_z^{ac}(x,y)$ information, the QOs at a single location or along 1D line scans are presented in the main text as the raw data of $B_z^{ac}$.

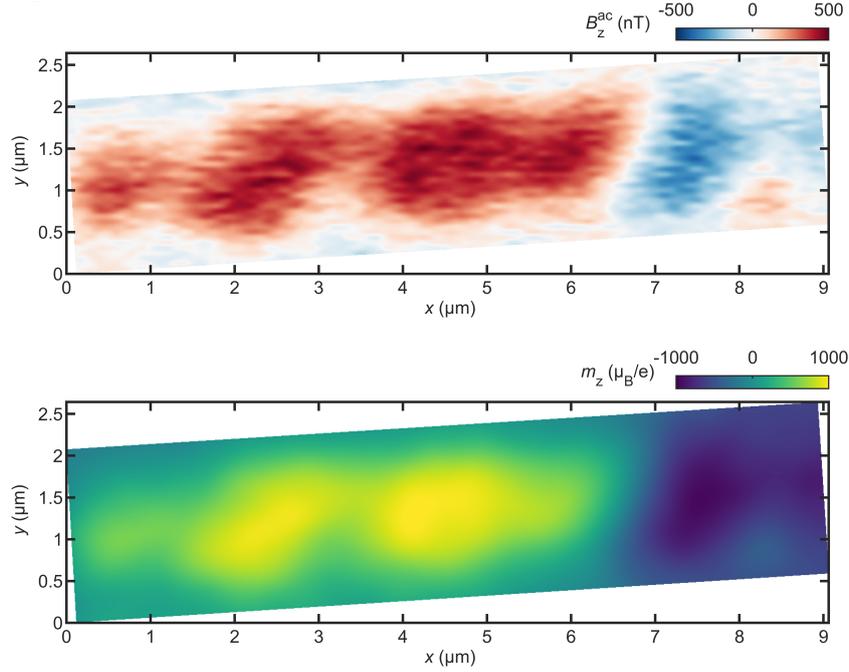

**Extended Data Fig. 2 | Reconstruction of the local magnetization from $B_z^{ac}(x,y)$. a**, Example of the measured $B_z^{ac}(x,y)$ at $n = 5.16 \times 10^{12}$ cm$^{-2}$, $B_a = 320$ mT, and $V_{bg}^{ac} = 10$ mV rms. **b**, Reconstructed differential magnetization $m_z = \partial M/\partial n$.

**Magnetic field and modulation amplitude dependence of QOs**

The measured signal $B_z^{ac} = n^{ac}(dB_z/dn)$ is proportional to the modulation amplitude of the carrier density $n^{ac}$ induced by $V_{bg}^{ac}$. It is therefore desirable to use large $n^{ac}$ to improve the signal-to-noise ratio. In order to resolve QOs, however, $n^{ac}$ has to be substantially smaller than the period of the oscillations $\Delta n$. Extended Data Figs. 3a-c show the QOs acquired at $B_a = 320$ mT using $V_{bg}^{ac} = 8$, 35, and 100 mV rms corresponding to $n^{ac}$ of $5.19 \times 10^9$ cm$^{-2}$, $2.27 \times 10^{10}$ cm$^{-2}$, and $6.49 \times 10^{11}$ cm$^{-2}$ rms. The four-fold degenerate BLG LLs have a period of $\Delta n = 4B_a/\phi_0 = 3.1 \times 10^{10}$ cm$^{-2}$. The lowest $V_{bg}^{ac} = 8$ mV rms was chosen to result in peak-to-peak value of $n^{ac}$ of $1.47 \times 10^{10}$ cm$^{-2}$, approximately equal to $\Delta n/2 = 1.55 \times 10^{10}$ cm$^{-2}$, which results in optimal signal-to-noise ratio for detecting the BLG LLs, albeit suppresses the measured $B_z^{ac}/n^{ac}$ ratio by a factor of $\pi/2$. A larger $n^{ac}$ washes out the QOs due to BLG LLs, leaving the MLG LLs clearly resolvable as demonstrated in Extended Data Figs. 3b,c. The largest $n^{ac}$ also allows clear observation of the paramagnetic response $\partial M/\partial \mu = C/\phi_0$ in the gap between the 0$^{th}$ and 1$^{st}$ MLG LLs dictated by the Chern number $C = 2$ on the electron side and $C = -2$ on the hole side (Extended Data Fig. 3c).

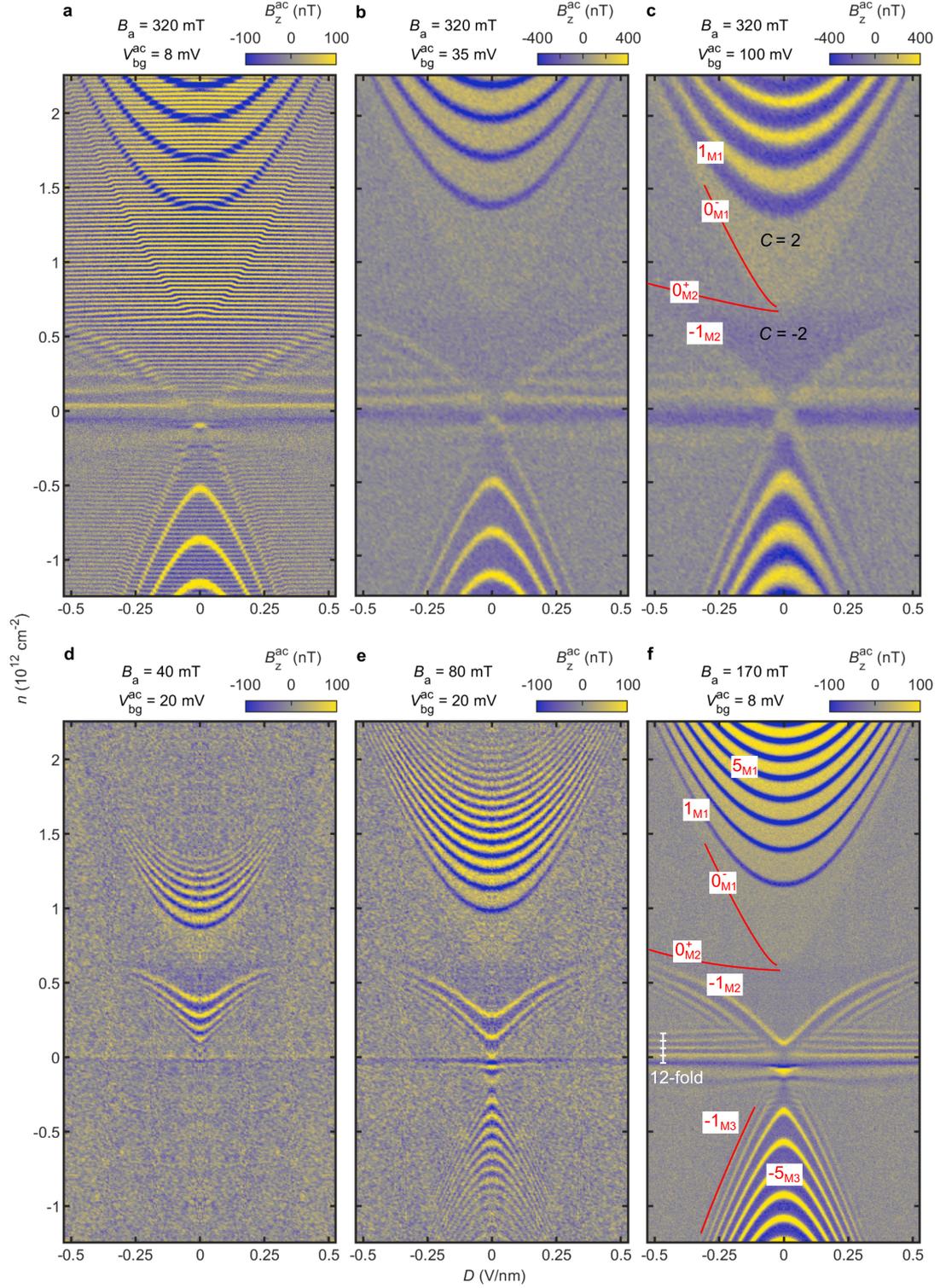

**Extended Data Fig. 3 | Comparison of $B_z^{ac}(n, D)$ at different magnetic fields and $V_{bg}^{ac}$. a,** The measured QOs at $B_a = 320$ mT and $V_{bg}^{ac} = 8$ mV rms. The induced peak-to-peak carrier density modulation, $n^{ac} = 1.47 \times 10^{10}$ cm$^{-2}$, is about half of the LL degeneracy $4B_a/\phi_0$ allowing clear resolution of the BLG and MLG LLs. **b,** QOs at $B_a = 320$ mT and $V_{bg}^{ac} = 35$ mV rms. The BLG LLs are washed out by the large carrier modulation while MLG LLs and the 12-fold degenerate LLs in the gullies are well resolved. **c,** At $V_{bg}^{ac} = 100$ mV rms the MLG QOs are smeared out. The paramagnetic response in the MLG LL gaps with Chern numbers $C = \pm 2$ are clearly visible as indicated. **d,** QOs at $B_a = 40$ mT and $V_{bg}^{ac} = 20$ mV rms. At this low field most of the LLs are smeared by the intrinsic LL broadening and only the lowest MLG LLs in sections M1 and M2 can be resolved at low displacement fields. **e,** Same as (d) at $B_a = 80$ mT. **f,** At $B_a = 170$ mT and $V_{bg}^{ac} = 8$ mV

rms the BLG LLs cannot be resolved, but all the MLG LLs and the 12-fold degenerate LLs in the gullies are very prominent.

Extended Data Figs. 3d-f show the QOs at $B_a = 40, 80$, and 170 mT. At these low fields the Dingle broadening greatly suppresses the QOs due to BLG LLs (see Extended Data Fig. 7), and reduces the visibility of the MLG LLs at large displacement fields due to the reduction in the gap energies. At 170 mT and $V_{bg}^{ac} = 8$ mV rms the M2 LLs and the 12-fold degenerate LLs in the gullies are resolved very clearly as seen in Extended Data Fig. 3f.

**Band structure calculations**

The band structure of ABA graphene was calculated within the tight-binding model following Ref. (*2*, *42*) based on SWMc parameterization (*40*). In the basis $\{A_1, B_1, A_2, B_2, A_3, B_3\}$, where $A_i$, $B_i$ are the two sublattice sites in the $i^{\text{th}}$ layer, the low energy effective Hamiltonian can be written as

$$H_0 = \begin{pmatrix} \Delta_1 + \Delta_2 & v_0\pi^\dagger & v_4\pi^\dagger & v_3\pi & \gamma_2/2 & 0 \\ v_0\pi & \delta + \Delta_1 + \Delta_2 & \gamma_1 & v_4\pi^\dagger & 0 & \gamma_5/2 \\ v_4\pi & \gamma_1 & \delta - 2\Delta_2 & v_0\pi^\dagger & v_4\pi & v_1 \\ v_3\pi^\dagger & v_4\pi & v_0\pi & -2\Delta_2 & v_3\pi^\dagger & v_4\pi \\ \gamma_2/2 & 0 & v_4\pi^\dagger & v_3\pi & -\Delta_1 + \Delta_2 & v_0\pi^\dagger \\ 0 & \gamma_5/2 & \gamma_1 & v_4\pi^\dagger & v_0\pi & \delta - \Delta_1 + \Delta_2 \end{pmatrix} \quad (1)$$

where $\Delta_1 = -e(U_1 - U_3)/2$ and $\Delta_2 = -e(U_1 - 2U_2 + U_3)/6$, with $U_i$ the potential of layer $i$. $\Delta_1$ is determined by the displacement field, while $\Delta_2$ describes the asymmetry of the electric field between the layers. The band velocities $v_i$ ($i$ = 0,3,4) are related to the tight-binding parameters $\gamma_i$ by $v_i\hbar = \frac{\sqrt{3}}{2}a_c\gamma_i$, where $a_c$ = 0.246 nm is the crystal constant of graphene, $\pi = \xi k_x + ik_y$, and $\xi$ is the valley index ($\xi = \pm 1$ for valley $K^+$ and $K^-$ respectively).

With a rotated basis $(A_1 - A_3)/\sqrt{(2)}, (B_1 - B_3)/\sqrt{(2)}, (A_1 + A_3)/\sqrt{(2)}, B_2, A_2, (B_1 + B_3)/\sqrt{(2)}$, the Hamiltonian can be rewritten as:

$$H_{TLG} = \begin{pmatrix} \Delta_2 - \frac{\gamma_2}{2} & v_0\pi^\dagger & \Delta_1 & 0 & 0 & 0 \\ v_0\pi & \Delta_2 + \delta - \frac{\gamma_5}{2} & 0 & 0 & 0 & \Delta_1 \\ \Delta_1 & 0 & \Delta_2 + \frac{\gamma_2}{2} & \sqrt{2}v_3\pi & -\sqrt{2}v_4\pi^\dagger & v_0\pi^\dagger \\ 0 & 0 & \sqrt{2}v_3\pi^\dagger & -2\Delta_2 & v_0\pi & -\sqrt{2}v_4\pi \\ 0 & 0 & -\sqrt{2}v_4\pi & v_0\pi^\dagger & \delta - 2\Delta_2 & \sqrt{2}\gamma_1 \\ 0 & \Delta_1 & v_0\pi & -\sqrt{2}v_4\pi^\dagger & \sqrt{2}\gamma_1 & \Delta_2 + \delta + \frac{\gamma_5}{2} \end{pmatrix} \quad (2)$$

For $\Delta_1 = 0$, the Hamiltonian can be block diagonalized into monolayer-like (MLG) and bilayer-like (BLG) blocks, i.e. $H_{TLG} = H_{MLG} \oplus H_{BLG}$. A finite displacement field hybridizes the two blocks.

In an external magnetic field, in the Landau gauge, the canonical momentum $\pi$ can be replaced by $\pi - eA$, where $A$ is the vector potential. $\pi$ obeys the commutation relation $[\pi_x, \pi_y] = -i/l_B$, where $l_B = \sqrt{(\hbar/eB)}$ is the magnetic length. As in the usual one-dimensional harmonic oscillator, in the basis of LL orbital $|n\rangle$, the matrix elements of $\pi$, $\pi^\dagger$ are given by:

$$
\begin{aligned}
K^+: \quad & \pi|n\rangle = \tfrac{i\hbar}{l_B}\sqrt{2(n+1)}|n+1\rangle \\
& \pi^\dagger|n\rangle = -\tfrac{i\hbar}{l_B}\sqrt{2n}|n-1\rangle \\
K^-: \quad & \pi|n\rangle = \tfrac{i\hbar}{l_B}\sqrt{2n}|n-1\rangle \\
& \pi^\dagger|n\rangle = -\tfrac{i\hbar}{l_B}\sqrt{2(n+1)}|n+1\rangle
\end{aligned}
\qquad (3)
$$

Therefore, the new Hamiltonian can be written in the basis of LL orbitals. Using matrix elements of $\pi$ and $\pi^\dagger$ operators, the momentum operators are replaced by an up (down) diagonal matrix of dimensions $\Lambda \times \Lambda$, where $\Lambda$ is the cutoff number for the infinite matrix, restricting the Hilbert space with indices $n \leq \Lambda$. All the other nonzero elements $C$ are substituted by $CI_\Lambda$, where $I_\Lambda$ is the identity matrix with dimensions $\Lambda \times \Lambda$. As our measurements were performed in low magnetic fields and high-index LLs are often involved, large cutoff was used so that it spans energy range significantly larger than in the experiment. We also removed 'false' LLs caused by imposing the cutoff, which usually have very large indices. In the simulations, $\Lambda$ was set to 400 for small carrier density range (Fig. 2) and to 800 for calculations over larger range (Fig. 3,4).

**Evolution of the band structure and Landau levels with displacement field**

Extended Data Fig. 4 shows the calculated BS of ABA graphene using the derived SWMc parameters and the evolution of the LLs with $D$ and $B_a$. At $D = 0$ ($\Delta_1 = 0$) there is essentially no hybridization between the MLG and BLG bands. All the LLs are valley (and spin) degenerate except for the 0$^{th}$ LLs of the MLG and BLG bands which are valley polarized due to the Berry curvature (Extended Data Fig. 4a). With increasing $\Delta_1$, the gaps of the MLG and BLG bands increase and the hybridization between the bands grows resulting in the formation of mini-Dirac cones (gullies) and in LL anticrossings (Extended Data Figs. 4b-d). At our highest accessible $\Delta_1 \cong 50$ meV the lowest LLs in the gullies are well isolated from the rest of the LLs as shown in Extended Data Figs. 4e,f. Since the BLG bandgap $\Delta_G^0$ is characterized by $C = 0$, it has no magnetization. The six-fold degenerated compressible 0$^{th}$ LLs $0_G^+$ and $0_G^-$ in the gullies also have no magnetization at low fields, $M = -\partial \epsilon / \partial B = 0$, due to their zero kinetic energy. As a result, zero magnetization is observed around the CNP over a width of $\delta n = 12 B_a / \phi_0$ in carrier density as indicated in Figs. 2c,f. The first paramagnetic signal appears when the Fermi level reaches the $C = \pm 6$ gaps $\Delta_G^1$ and $\Delta_G^{-1}$ between the 0$^{th}$ and the 1$^{st}$ gully LLs as marked in Fig. 2f. At elevated magnetic fields the six-fold gully degeneracy of the 0$^{th}$ LLs is partially lifted (*4*, *57*).

**Reconstruction of band structure parameters**

A number of experimental studies (*3, 4, 43, 44, 46, 55, 56, 58, 68*) have investigated the tight-binding parameters of ABA graphene as summarized in Table 1. The high resolution of our data and the fine features attained at low magnetic fields allow high precision reconstruction of SWMc parameters as follows. We set $\gamma_0$ to the standard literature value of 3100 meV and fit the remaining seven parameters. Extended Data Fig. 5 shows the effect of the individual parameters on the BS.

In the absence of displacement field, $\Delta_1 = 0$, the MLG band is affected only by $\gamma_0$, $\gamma_2$, $\gamma_5$ and $\delta$, with Fermi velocity $v_0$ given by $\gamma_0$ and the gap at the Dirac point by $E_g^0 = \delta + \frac{\gamma_2 - \gamma_5}{2}$. The BLG band is strongly dependent on $\gamma_0$, $\gamma_1$ and $\gamma_3$, weakly dependent on $\gamma_4$, and essentially independent of $\gamma_5$ and $\delta$. The BLG gap size is mainly governed by $\gamma_2$ and $\Delta_2$. The relative energy shift between the MLG and BLG bands is mainly governed by $\gamma_2$.

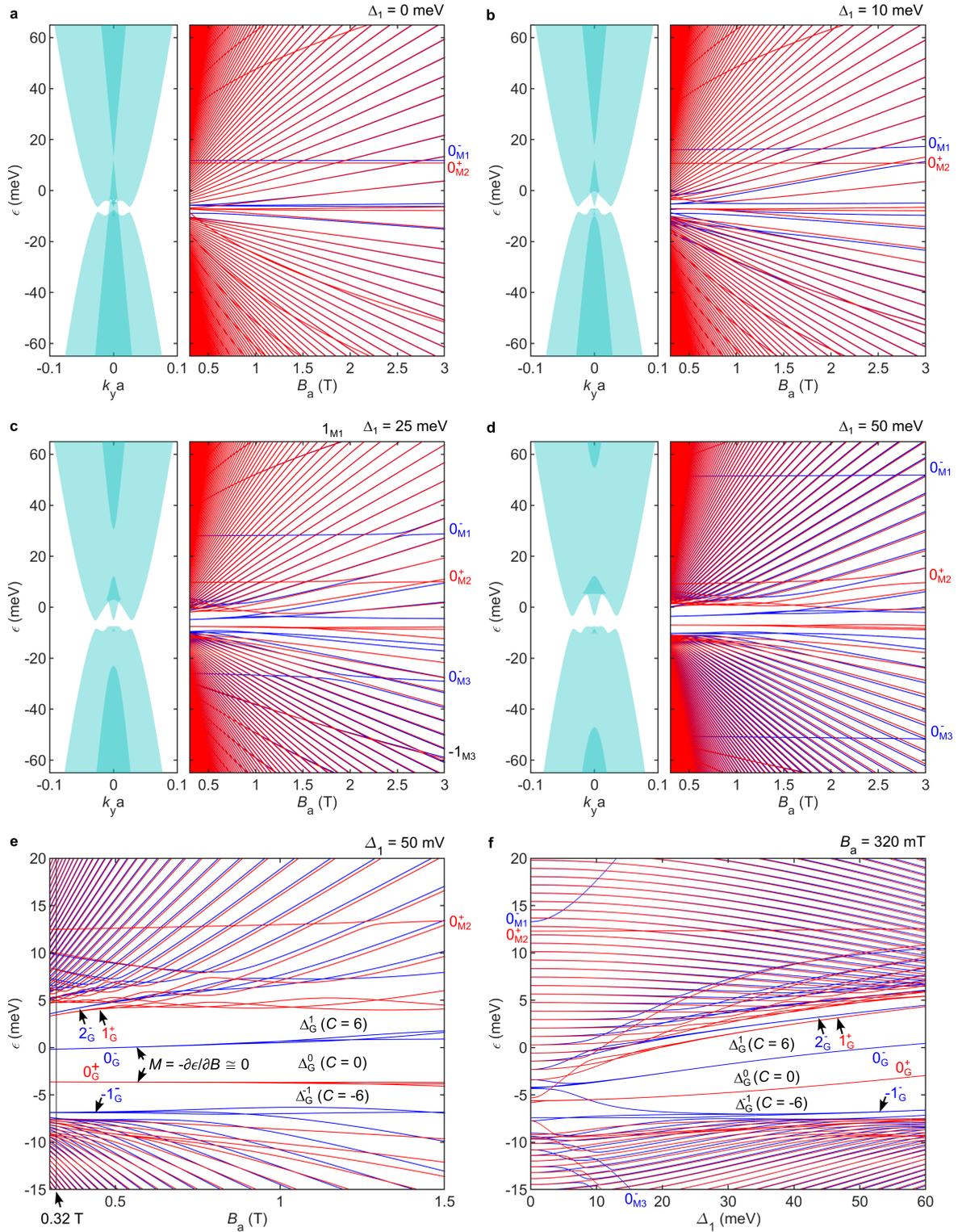

**Extended Data Fig. 4 | Evolution of ABA graphene band structure with displacement field. a-d**, Projected 3D band structure of ABA graphene (left panels) and the corresponding evolution of LLs with $B_a$ at different values of potential difference $\Delta_1 = 0, 10, 25,$ and $50$ meV between the top and bottom graphene layers (right panels). Red and blue lines denote the LLs in $K^+$ and $K^-$ valleys respectively. The tight-binding parameters used for the calculations are given in Table 1 (bottom row). With increasing $\Delta_1$, the MLG and BLG bandgaps grow and band hybridization is enhanced forming mini-Dirac cones (gullies) near CNP. **e**, Zoomed-in view of the evolution of LLs with $B_a$ at $\Delta_1 = 50$ meV. At low fields the $0^{th}$ LLs $0_G^+$ and $0_G^-$ are six-fold degenerate (including spin) and have vanishing magnetization. The six-fold degeneracy of the gully LLs is partially lifted at high $B_a$. The Chern numbers $C$ in the large gaps are indicated. **f**, Evolution of LLs with $\Delta_1$ at $B_a = 320$ mT.

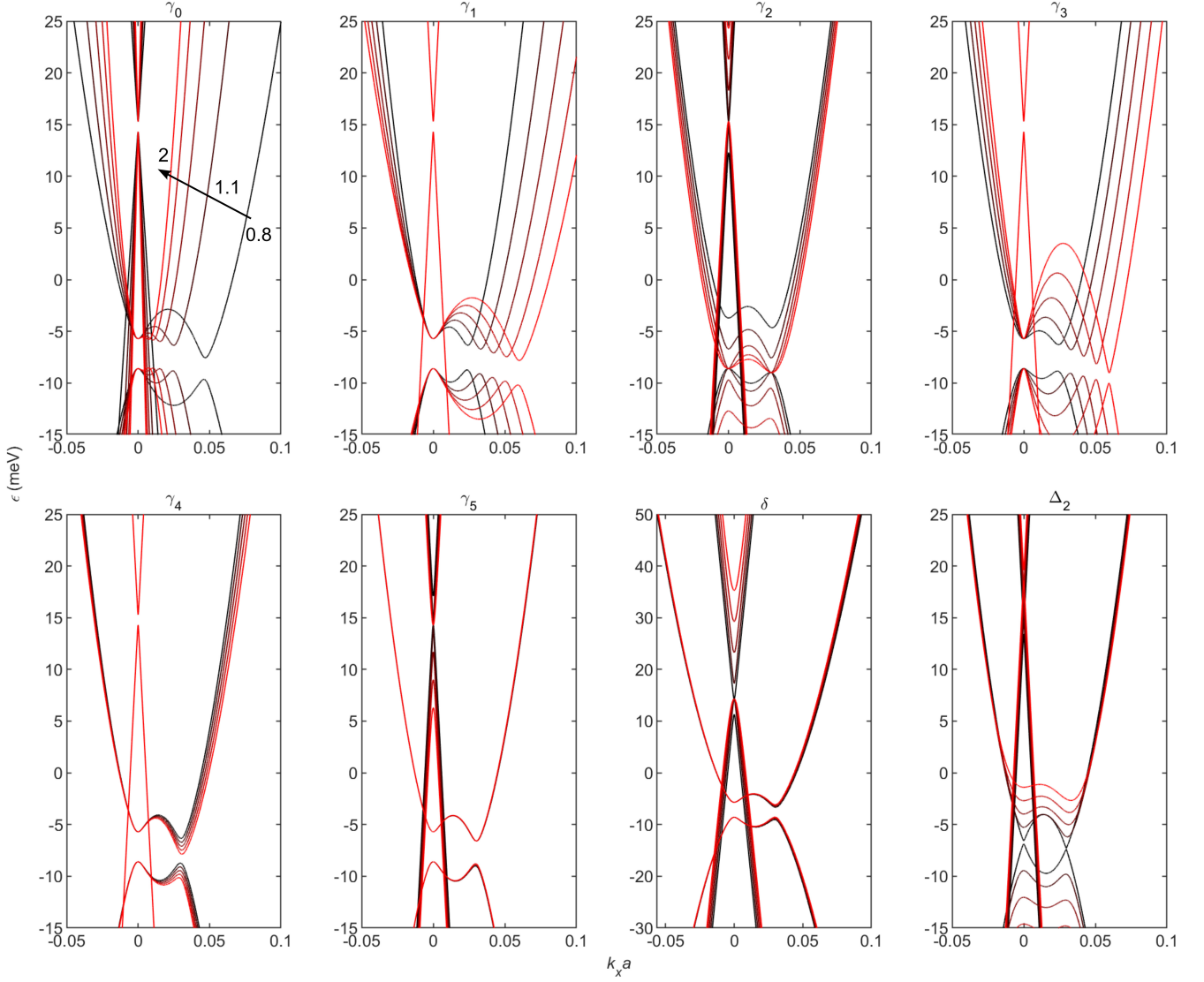

**Extended Data Fig. 5 | The dependence of the band structure on the SWMc parameters $\gamma_0$ to $\gamma_5$, $\delta$, and $\Delta_2$ at $\Delta_1 = 0$.** The panels show the effect of individual parameters on the BS, calculated using the parameters of Ref. (*46*) (see Table 1) and multiplied by a factor 0.8, 1.1, 1.4, 1.7 and 2 denoted by the different colors from black to red.

The dependence of the measured QOs on $n$ and $D$ at low $B_a$ provides a very sensitive tool for determining the SWMc parameters. The following attributes are particularly informative:

(i) The number of BLG LLs between the adjacent MLG LLs.
(ii) The relative energy shift between MLG and BLG bands.
(iii) LL anticrossings in the gullies.
(iv) The gap size of MLG band.

Attribute (i) is determined by the DOS ratio of the two bands. For a given $\gamma_0$, the relative DOS is predominantly governed by $\gamma_1$. By adjusting $\gamma_1$ to fit the relative number of BLG and MLG LLs along with optimization of other parameters we attain $\gamma_1 = 370 \pm 10$ meV.

Attribute (ii) is then used to determine $\gamma_2$. The energies of the band extrema and hence the relative position of the 0$^{th}$ LLs can be calculated analytically. In particular, from Eq. (2) for $\Delta_1 = 0$, the $0^-_{M1}$ LL at the bottom of M1 band is positioned at energy $\Delta_2 - \gamma_2/2$, while the top of BLG valence band is at $\Delta_2 + \gamma_2/2$. Thus, the relative position between MLG and BLG bands is determined by $\gamma_2$ and $\Delta_2$. Since the LL spectrum is quite sensitive to $\Delta_2$, $\gamma_2$ is determined first. We use the relative position between $-1_{M3}$ and the nearby BLG LLs to fit $\gamma_2$, and we get $\gamma_2 = -19 \pm 0.5$ meV.

Attribute (iii) is governed by $\gamma_3$, which induces trigonal warping of the BLG bands. As shown in Extended Data Fig. 8, this results in the anticrossings between the BLG LLs and MLG $0^-_{M3}$ and $-1^+_{M3}$ LLs. From fitting to the experimental data we attain $\gamma_3 = 315 \pm 10$ meV.

Attributes (ii) and (iv) are used to derive $\delta$ and $\gamma_5$. From Eq. (2), the MLG band gap at $D$ = 0 V/nm is $E_g^0 = \delta + (\gamma_2 - \gamma_5)/2$, while the gap center is located at $2\Delta_2 + \delta - (\gamma_2 + \gamma_5)/2$. In our experiment data, one BLG LL fits within the MLG gap and 20 BLG LLs reside between $0^-_{M1}$ and $-1_{M3}$, from which we attain $\delta = 18.5 \pm 0.5$ meV and $\gamma_5 = 20 \pm 0.5$ meV. Note that $E_g^0$ can be either positive or negative. We find that $E_g^0$ is negative, which means that the 0$^{th}$ $K^-$ LL ($0^-_{M1}$) resides at the bottom of M1 band and the 0$^{th}$ $K^+$ LL ($0^+_{M2}$) is at the top of M2. In this case the Dirac gap $E_g$ increases with $\Delta_1$ and the $0^-_{M1}$ and the $0^+_{M2}$ LLs spread apart with the displacement field as shown in Extended Data Fig. 4f, consistent with experimental data in Figs. 2c,e and calculations in Figs. 2d. If $E_g^0$ is positive, the 0$^{th}$ $K^-$ LL will reside at the top of M2 while the 0$^{th}$ $K^+$ LL will be at the bottom of M1. In this case, upon increasing $D$, the Dirac gap closes and then reopens with crossing of the two 0$^{th}$ LLs, such that $E_g$ is always negative at high $D$ with 0$^{th}$ $K^-$ LL at the bottom of M1. Table 1 shows that the value and the sign of $E_g^0$ varies significantly in the literature. Note, however, that only in Datta(*46*) and in the present work the Dirac gap is reported directly. For the rest of the references the $E_g^0$ values presented in the table are calculated from the reported values of $\delta$, $\gamma_2$, and $\gamma_5$.

$\Delta_2$ mainly affects the gap of the BLG bands and as $-1_{M3}$ resides closely to the BLG band gap, we use the number of BLG LLs between $-1_{M3}$ and $-2_{M3}$ to fit $\Delta_2$ and get $\Delta_2 = 3.8 \pm 0.05$ meV. $\gamma_4$ plays the most negligible role, slightly adjusting the shape of the BLG bands. The fitting procedure is to choose these parameters such that the inaccuracy of the number of BLG LLs between any pair of MLG LLs is no more than 1. By optimizing all parameters for best fit to the experimental data we derive $\gamma_4 = 140 \pm 15$ meV, as summarized in Table 1.

**Table 1 | Comparison of SWMc parameters in different works.** The parameters are in units of meV.

| Reference | $\gamma_0$ | $\gamma_1$ | $\gamma_2$ | $\gamma_3$ | $\gamma_4$ | $\gamma_5$ | $\delta$ | $\Delta_2$ | $E_g^0$ |
|---|---|---|---|---|---|---|---|---|---|
| Taychatanapat (*3*) | 3160 | 390 | -28 | 315 | 41 | 50 | 46 | Nan | 7 |
| Shimazaki (*55*) | 3100 | 390 | -28 | 315 | 41 | 50 | 34 | 0 | -5 |
| Stepanov (*58*) | 3230 | 310 | -32 | 300 | 40 | 10 | 27 | 1.8 | 6 |
| Campos (*43*) | 3100 | 390 | -18 ± 2 | 315 | 90 ± 50 | 10 ± 5 | 15 ± 3 | >2.7 | 1 |
| Datta (*56*) | 3100 | 390 | -28 | Nan | Nan | 10 | 21 | Nan | 2 |
| Asakawa (*68*) | 3230 | 390 | -23.7 | 315 | 43.8 | 6 | 14.3 | Nan | -0.55 |
| Datta (*46*) | 3100 | 390 | -20 | 315 | 120 | 18 | 20 | 4.3 | 1 |
| Zibrov (*4*) | 3100 | 380 | -20 | 290 | 141 | 50 | 35.5 | 3.5 | 0.5 |
| Che (*44*) (suspended) | 3100 | 355 | -41 | 315 | 150 | 40 | 47 | 1 | 6.5 |
| Che (*44*) (*h*BN-aligned) | 3100 | 355 | -12.5 | 315 | 150 | 40 | 31.5 | 5.8 | 5.25 |
| This work | 3100 | 370 ± 10 | -19 ± 0.5 | 315 ± 10 | 140 ± 15 | 20 ± 0.5 | 18.5 ± 0.5 | 3.8 ± 0.05 | -1.0 |

**Orbital magnetization calculations**

Oscillations in orbital magnetization $M$ due to LLs can be calculated analytically for either parabolic or Dirac bands as shown previously (*69*). However, there is no analytical expression for the LL spectrum in ABA graphene, and therefore the magnetization oscillations have to be calculated numerically. We follow the method described in (*14*) to derive the magnetization $M(n)$ and then calculate its derivative $\partial M/\partial n$.

We first consider the case with zero LL broadening. For an arbitrary LL spectrum $E_i$ with degeneracy $D_i$ ($i$ is the Landau level index), the density of states $N_0(\epsilon)$ of the system is:

$$N_0(\epsilon) = \sum_i D_i\, \delta(\epsilon - E_i).$$

$E_i$ describes spin-degenerate LLs from both valleys with degeneracy $D_i = 2\frac{eB}{h}$. The grand thermodynamic potential $\Omega_0(\mu, B)$ is then given by:

$$\Omega_0 = -kT \int_{-\infty}^{\infty} N_0(\epsilon) \ln\left(1 + e^{[(\mu - \epsilon)/kT]}\right) d\epsilon.$$

Here $k$ is the Boltzmann constant, $T$ is the temperature and $\mu$ is the chemical potential.

Now we consider LL broadening of width $\Gamma$ (Dingle parameter) with a Lorentzian form

$$L(\epsilon) = \frac{1}{\pi}\left(\frac{\Gamma}{\epsilon^2 + \Gamma^2}\right).$$

The DOS and the grand potential are then described by

$$N(\epsilon) = \sum_i D_i L(\epsilon - E_i),$$

$$\Omega = -kT \int_{-\infty}^{\infty} N(\epsilon) \ln\left(1 + e^{[(\mu - \epsilon)/kT]}\right) d\epsilon = -kT \int_{-\infty}^{\infty} \sum_i D_i\, L(\epsilon - E_i) \ln\left(1 + e^{[(\mu - \epsilon)/kT]}\right) d\epsilon$$

Then $M$ is given by

$$M = -\frac{\partial \Omega}{\partial B} = kT \int_{-\infty}^{\infty} \sum_i \left(\frac{\partial D_i}{\partial B} L(\epsilon - E_i) - D_i L'(\epsilon - E_i)\frac{\partial E_i}{\partial B}\right) \ln\left(1 + e^{[(\mu - \epsilon)/kT]}\right) d\epsilon,$$

where $L'(\epsilon) = \partial L(\epsilon)/\partial \epsilon$. In the zero-temperature limit ($T\to 0$), $M$ can be simplified:

$$M = \int_{-\infty}^{\mu} \sum_i \left(\frac{\partial D_i}{\partial B} L(\epsilon - E_i) - D_i L'(\epsilon - E_i)\frac{\partial E_i}{\partial B}\right)(\mu - \epsilon) d\epsilon.$$

To compare with our experiment, we need to calculate

$$\frac{\partial M}{\partial n}(n) = \frac{\partial M}{\partial \mu}(n)\frac{\partial \mu}{\partial n}(n),$$

where $\frac{\partial \mu}{\partial n}(n)$ is the inverse of the density of states as a function of the carrier density and $n(\mu) = \int_{-\infty}^{\mu} N(\epsilon) d\epsilon$.

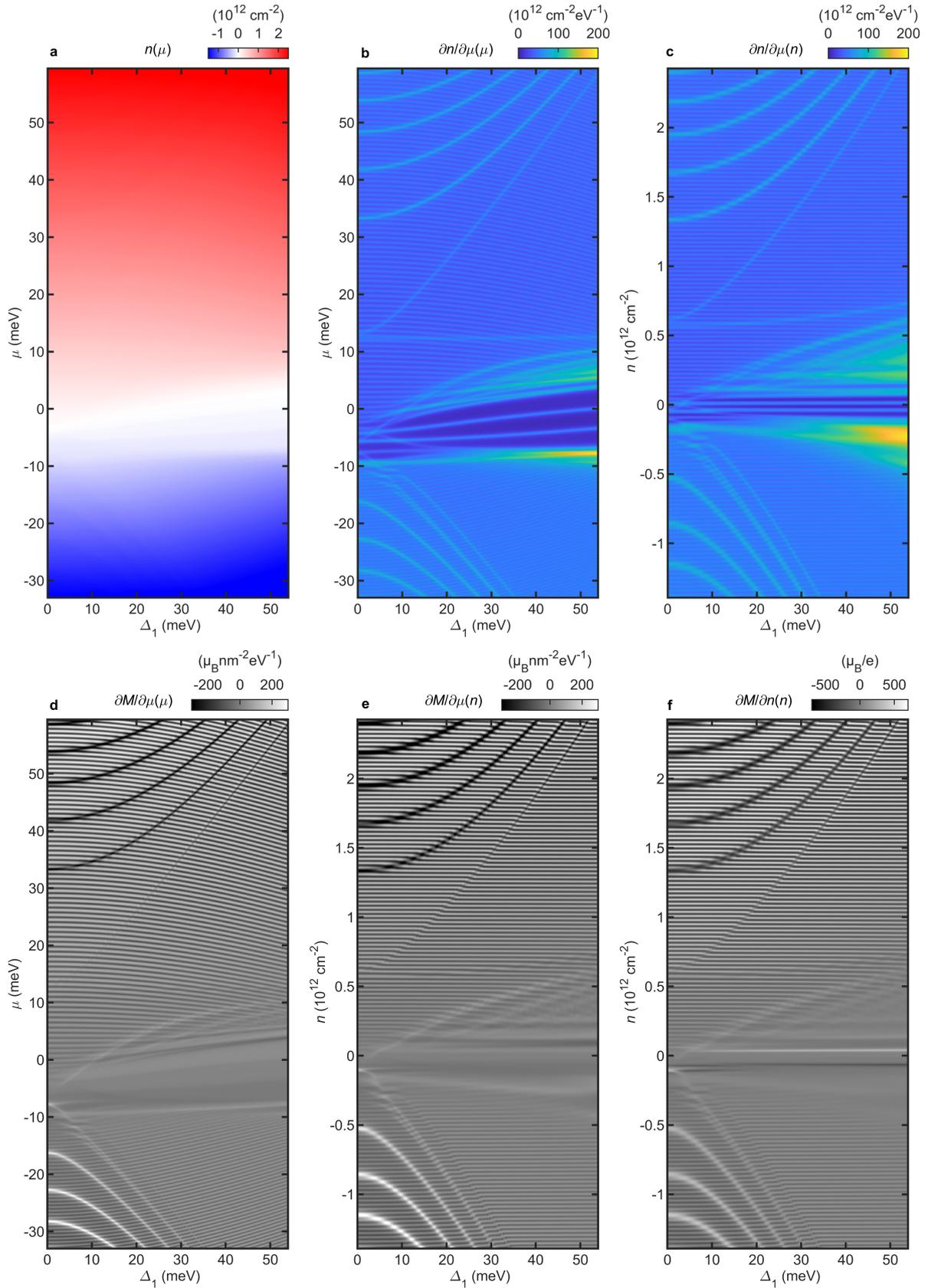

**Extended Data Fig. 6 | Calculations of the orbital magnetization. a,** Calculated carrier density $n$ as a function of chemical potential $\mu$ and displacement-field-induced potential difference $\Delta_1$. **b,** $\partial n/\partial \mu$ vs. $\mu$ and $\Delta_1$. **c,** $\partial n/\partial \mu$ vs. $n$ and $\Delta_1$. **d,** Calculated differential magnetization $\partial M/\partial \mu$ vs. $\mu$. **e-f,** Calculated $\partial M/\partial \mu$ (e) and differential magnetization $m_z = \partial M/\partial n$ (f) vs. $n$. $B_a = $ 320 mT and $\Gamma = 0.3$ meV in all the calculations.

Extended Data Fig. 6 shows the calculated $n(\mu)$, $\frac{\partial n}{\partial \mu}(\mu)$, $\frac{\partial n}{\partial \mu}(n)$, $\frac{\partial M}{\partial \mu}(\mu)$, $\frac{\partial M}{\partial \mu}(n)$, and $\frac{\partial M}{\partial n}(n)$ vs. $\Delta_1$ at $B_a =$ 320 mT using the derived SWMc parameters and Dingle broadening $\Gamma = 0.3$ meV. The modulation in DOS, $\partial n/\partial \mu$, is well resolved in Extended Data Figs. 6b,c, but it is relatively small due to the LL broadening, except near CNP where large gaps with vanishing DOS open between the lowest LLs in the gullies at elevated $\Delta_1$.

The calculated $\partial M/\partial \mu$ vs. $\mu$ in Extended Data Figs. 6d shows that the crossing of the MLG and BLG LLs does not cause any phase shift. In contrast, in $\partial M/\partial \mu$ vs. $n$ in Extended Data Figs. 6e, the BLG LLs show a $2\pi$ shift upon crossing the four-fold degenerate MLG LLs and a $\pi$ shift upon crossing the two-fold degenerate 0$^{th}$ LLs. This arises from the fact the filling a MLG LL delays filling the next BLG LL vs. total $n$, but not vs. $\mu$. Since the DOS modulation $\frac{\partial n}{\partial \mu}(n)$ is quite small, $\frac{\partial M}{\partial n}(n)$ in Extended Data Figs. 6c looks very similar to $\frac{\partial M}{\partial \mu}(n)$ except near CNP.

**Derivation of the Dingle parameter**

At our low $B_a$ the intrinsic LL broadening strongly suppresses the amplitude of the measured QOs. Extended Data Fig. 7 shows the calculated QOs for various Dingle parameters $\Gamma = 0.2$ to 0.8 meV. Since the energy spacing of the BLG LLs in the conduction band is about 1 meV, the amplitude of their QOs is suppressed by about two orders of magnitude over this range of $\Gamma$, while in the valence band, where the LL gaps are about 0.6 meV, the QOs are completely quenched with the higher $\Gamma$. In contrast, the amplitude of the QOs of the MLG LLs, which have an order of magnitude larger gaps at low carrier densities, is much less affected by these $\Gamma$ values. As a result, the relative amplitude of the MLG and BLG QOs is strongly dependent on $\Gamma$, allowing its accurate determination. By fitting to the experimental data in Fig. 2, we attain $\Gamma = 0.3 \pm 0.05$ meV, which also provides a very good agreement in quantitative comparison between the amplitudes of the measured $B_z^{ac}$ and the calculated $m_z$ taking into account the 2D magnetization reconstruction.

Note that the finite $n^{ac}$ modulation by $V_{bg}^{ac}$ also causes a suppression of the apparent amplitude of the QOs. It can be shown that if the peak-to-peak value of the carrier density modulation is less than half of the LL degeneracy, $n^{ac} < 2B_a/\phi_0$, which is the case in our high-resolution measurements, the suppression is less than a factor of $\pi/2$. For larger $n^{ac}$ the visibility is suppressed rapidly as shown in Extended Data Fig. 3. In particular, in Figs. 3 and 4 we have intentionally used larger $n^{ac}$ in order to suppress the QOs due to BLG LLs and to improve the signal to noise ratio for detections of the MLG LLs. Since this type of suppression of the apparent amplitude of QOs is harder to simulate in our BS calculations, we have used $\Gamma = 0.3$ meV for the calculations presented in all the figures except for Figs. 3 and 4, where $\Gamma = 0.6$ meV was used instead for suppression of the BLG QOs artificially. This larger $\Gamma$ does not affect appreciably the shape of the calculated MLG QOs but reduces their amplitude.

Our derived $\Gamma = 0.3$ meV with corresponding local quantum scattering time $\tau_q = \hbar/2\Gamma \cong 1$ ps, is about four times lower than the value reported based on global SdH oscillations (*46*). This is consistent with the observation that the lowest magnetic field for detection of QOs in our local dHvA measurements is substantially lower than what is required for detection of the SdH oscillations (Extended Data Fig. 1). The large $\Gamma$ reported based on SdH oscillations is likely due to sample inhomogeneity, such as charge disorder and the pseudomagnetic fields ($B_S$).

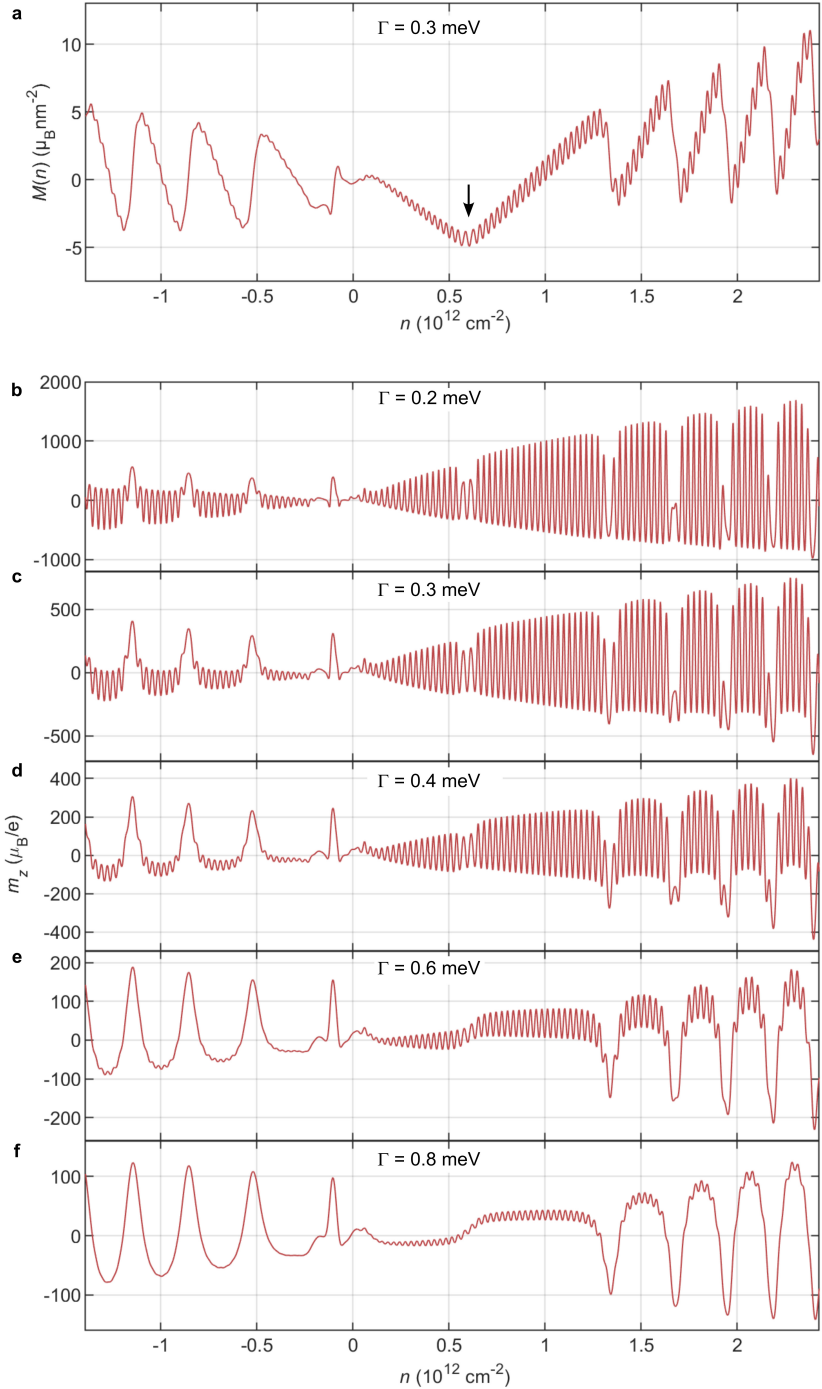

**Extended Data Fig. 7 | Suppression of QOs by LL broadening**. **a,** Calculated $M(n)$ at $D$ = 0 V/nm, $B_a$ = 320 mT, and Dingle parameter $\Gamma = 0.3$ meV. The V-shaped dip in $M$ (black arrow) corresponds to the McClure peak at the Dirac point of the MLG band (14). **b-f,** Calculated differential magnetization $m_z = \partial M/\partial n$ for different $\Gamma = 0.2$ to 0.8 meV. The QOs due to BLG LLs with small energy gaps are suppressed much stronger by $\Gamma$ than the MLG LLs with large gaps.

**LL anticrossings**

The hybridization between the BLG and MLG bands upon increasing $\Delta_1$ with the displacement field gives rise to partial lifting of valley degeneracy of the LLs. This effect is particularly pronounced near the top of the BLG valence band at intermediate values of $\Delta_1$ as shown in Extended Data Figs. 8c,d. Here, when MLG and BLG LLs in the same valley intersect, the strong band hybridization and non-vanishing $\gamma_3$ leads to avoided crossing between the LLs as marked by the open symbols. Interestingly, the anticrossing occurs between the MLG LLs and every third BLG LL. Our derived SWMc parameters provide an excellent fit to the experimentally observed anticrossings as demonstrated in Extended Data Figs. 8a,b. Moreover, the strong hybridization lifts the valley degeneracy of the 1st MLG LL in the M3 sector as shown by the pronounced splitting between $-1^-_{M3}$ and $-1^+_{M3}$ in Extended Data Figs. 8b-d. This novel splitting is clearly resolved experimentally in Extended Data Figs. 8a.

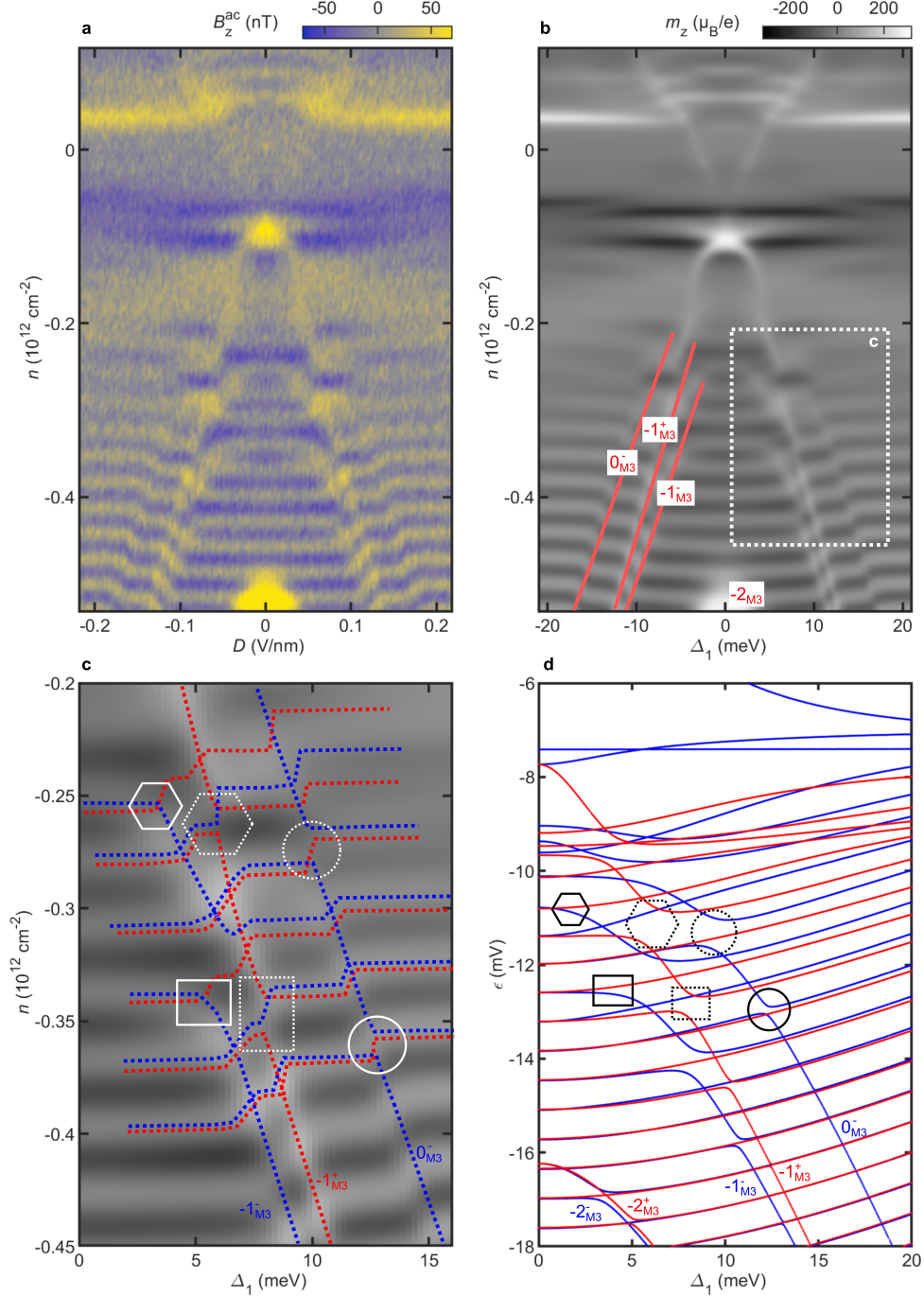

**Extended Data Fig. 8 | Landau level anticrossings**. **a,** Measured QOs in $B_z^{ac}$ vs. $n$ and $D$ at $B_a = 320$ mT near the top of the BLG valence band reproduced from Fig. 2f. **b,** Calculated $m_z$ using the derived SWMc parameters providing an excellent fit to the experimental data. **c,** A zoom-in of the region marked in (b) with overlaid schematic LLs. **d,** Calculated LLs in the $K^-$ (blue) and $K^+$ (red) valleys. The open symbols denote the points of valley splitting and LL anticrossings between the MLG and BLG LLs. The splitting of $-1_{M3}$ LL into valley polarized $-1^-_{M3}$ and $-1^+_{M3}$ LLs is clearly resolved in calculations and in the experimental data.

## Interference of BLG Landau levels

The interference of the LLs can be observed also in the BLG bands at the same locations where it is present in the MLG bands. Extended Data Fig. 9 shows the QOs acquired at site B as in Figs. 3h and 4e, but using lower $V_{bg}^{ac} = 8$ mV rms that allows resolving the BLG LLs. The beating nodes at around 0.5 and $1.8 \times 10^{12}$ cm$^{-2}$ are clearly seen (Extended Data Fig. 9b), which can be well reproduced by the simulations using $B_S = 4.2$ mT (Extended Data Fig. 9c).

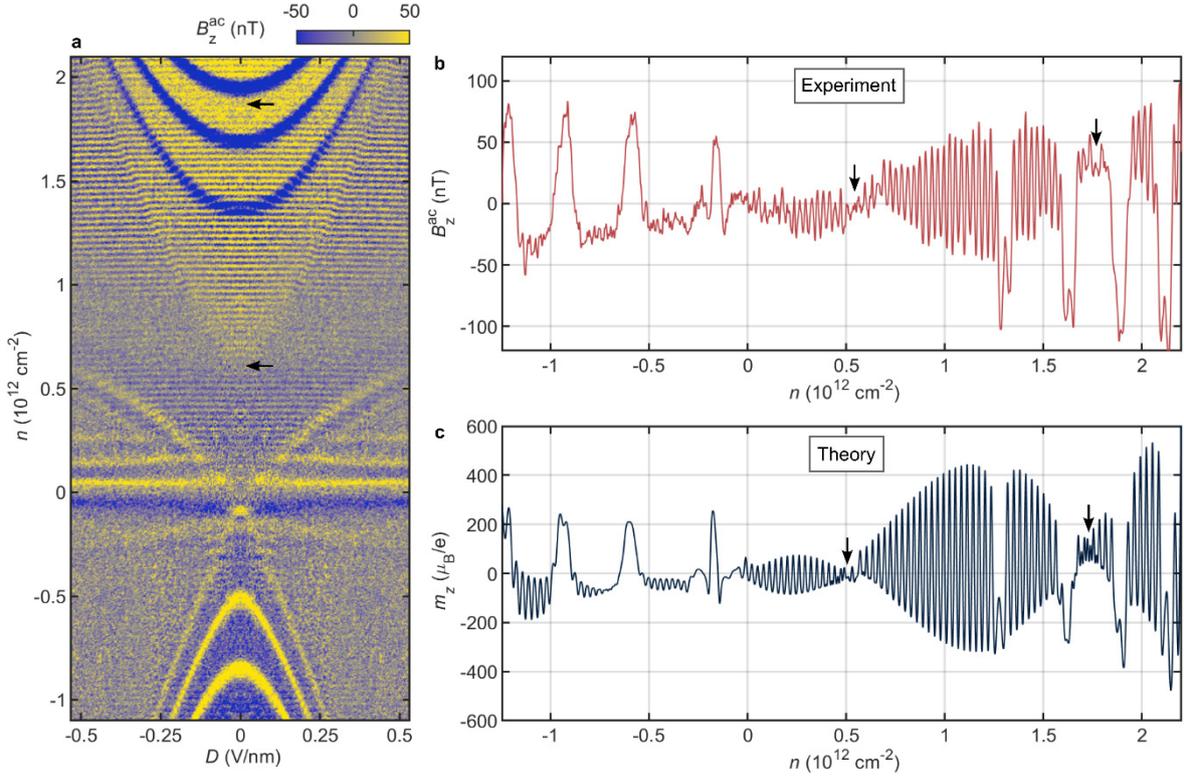

**Extended Data Fig. 9 | Beating of the BLG LLs. a**, QOs vs. $n$ and $D$ at $B_a = 320$ mT acquired at site B in Fig. 3b using $V_{bg}^{ac} = 8$ mV rms. **b,** $B_z^{ac}$ vs. $n$ profile at $D = 0$ V/nm. **c,** Calculated QOs with $B_S = 4.2$ mT. The black arrows indicate the positions of the beating nodes.

## Resolution of the pseudomagnetic field measurement by Landau level interference

The minimal PMF that can be measured using the interference method is determined by the highest accessible LL index of the beating node $N_b^1$. At $B_a = 320$ mT within the accessible range of $n$ the highest MLG LL index in ABA graphene is $\pm 70$, and hence the minimal $B_S = B_a/(4N_b^1) = 1.14$ mT. For comparison, the lowest PMF that has been recently resolved by STM is $B_S \cong 0.5$ T (70).

## Discussion of possible alternative mechanisms of interference of QOs

We consider below several additional possible mechanisms that can alter the BS and induce degeneracy lifting, which may lead to interference of the LLs, and show that they are incompatible with the experimental data.

***Band shifting***. Spin-orbit coupling (SOC) as well as Zeeman effect at elevated fields can lift flavor degeneracy producing an energy shift between the bands of opposite spin or valley. Both the intrinsic SOC in graphene and the Zeeman contributions at our low magnetic fields result in a negligible energy shift of the order of µeV (71, 72), which cannot account for the experimental data. Nevertheless, we explore whether a generic rigid shift between bands can reproduce the revealed LL interference pattern. In Fig. 3h the first node in the interference of the MLG LLs occurs at index $N \cong 19$. The corresponding LL energy gap is $\Delta E_N = E_{N+1} - $ $E_N = \sqrt{2e\hbar v_F^2 B_a}(\sqrt{N+1} - \sqrt{N}) \cong 2.5$ meV. For the destructive interference, the LLs of the two bands have to be out of phase, namely shifted by $\delta E_N \cong 1.25$ meV. Extended Data Fig. 10a shows the BS with a rigid shift of 1.25 meV between the $K^+$ and $K^-$ bands with the corresponding calculated QOs presented in Extended Data Fig. 10b. The main resulting feature is that the MLG LLs are split into two which is markedly different from the experimental QOs. This clearly points out that in order to reproduce the observed QOs, the energy shift $\delta E_n$ between the interfering LLs has to grow with the LL index rather than being constant or decreasing with $N$. This is indeed the behavior in the case of PMF, where $\delta E_N = \sqrt{2e\hbar v_F^2 N}(\sqrt{B_a + B_S} - \sqrt{B_a - B_S}) \cong B_S\sqrt{2e\hbar v_F^2 N/B_a}$ grows as $\sqrt{N}$.

***Staggered substrate potential***. The possible alignment between the hBN and ABA graphene can cause an onsite potential difference between the A and B sublattices. Here we consider the simplest situation where one of the graphene layers (bottom) is aligned with the hBN giving rise to a staggered substrate potential. In this case, the Hamiltonian can be written in the basis of $\{A_1, B_1, A_2, B_2, A_3, B_3\}$ as:

$$H_0 = \begin{pmatrix} \Delta_1 + \Delta_2 & v_0\pi^\dagger & v_4\pi^\dagger & v_3\pi & \gamma_2/2 & 0 \\ v_0\pi & \delta + \Delta_1 + \Delta_2 & \gamma_1 & v_4\pi^\dagger & 0 & \gamma_5/2 \\ v_4\pi & \gamma_1 & \delta - 2\Delta_2 & v_0\pi^\dagger & v_4\pi & v_1 \\ v_3\pi^\dagger & v_4\pi & v_0\pi & -2\Delta_2 & v_3\pi^\dagger & v_4\pi \\ \gamma_2/2 & 0 & v_4\pi^\dagger & v_3\pi & -\Delta_1 + \Delta_2 + \delta_{A_3} & v_0\pi^\dagger \\ 0 & \gamma_5/2 & \gamma_1 & v_4\pi^\dagger & v_0\pi & \delta - \Delta_1 + \Delta_2 + \delta_{B_3} \end{pmatrix}$$

For concreteness, we choose $\delta_{A3} = 2$ meV and $\delta_{B3} = -2$ meV. The resulted BS is shown in Extended Data Fig. 10c (red) in comparison to the original BS (black). The staggered substrate potential increases the gaps of the MLG and BLG bands but does not lift the valley degeneracy and therefore does not lead to beating. Extended Data Fig. 10d presents the calculated QOs showing no LL beating.

***Kekulé distortion***. Kekulé distortions are the bond density waves that have been observed in graphene epitaxially grown on copper (*73*) or in the presence of strain (*74*). In contrast to the O-type Kekulé distortion which opens a gap at the Dirac point, we find that the Y-type (*75*) can result in LL interference. The Y-shaped modulation of the bond strength, parametrized by the hopping parameters $\gamma_0$ and $\gamma_0'$ (Extended Data Fig. 10e), gives rise to valley-momentum locking and to inequivalent Fermi velocities for both the MLG and BLG bands. Hence, it lifts the valley degeneracy of the LLs resulting in chiral symmetry breaking. Within the SWMc model, $\gamma_0$ is the sole parameter that controls the Fermi velocity $v_F$ of the MLG band ($v_F = \frac{\sqrt{3}}{2}\frac{a\gamma_0}{\hbar}$). The energy difference between the LLs from the two valleys with the same index $N$ is $\delta E_N = \sqrt{2e\hbar NB_a}\Delta v_F$, where $\Delta v_F = \frac{\sqrt{3}}{2}\frac{a}{\hbar}(\gamma_0 - \gamma_0')$. The first beating node appears when $\delta E_n$ is equal to half of the gap size: $\sqrt{2e\hbar NB_a}\Delta v_F = \sqrt{e\hbar v_F^2 B_a/2N}/2$, which yields $N_b^1 = v_F/(4\Delta v_F)$ as shown in Extended Data Fig. 10f. In Fig. 3h $N_b^1 = 19$, which would correspond to a very weak Kekulé distortion with $\Delta v_F/v_F = 1.4 \times 10^{-2}$. Note, however, that the Kekulé distortion results in $N_b^1$ that is independent of $B_a$ as corroborated by the calculated QOs for $B_a = 320$ and 170 mT in Extended Data Figs. 10h,i. The reason being that the LLs shift in the same proportion in the two valleys with $B_a$. This is in sharp contrast to beating due to PMF for which $N_b^1 = B_a/(4B_S)$ is proportional to $B_a$. The experimental data points in Extended Data Fig. 10g (circles) are clearly consistent with PMF and incompatible with the Kekulé distortion.

***Disorder in band structure parameters.*** The BS can vary in space due to various types of disorder. Focusing on the Dirac bands, for example, the energy of the Dirac point or $v_F$ could be position dependent without breaking the valley symmetry. If the parameters change gradually in space on lengths scale larger than our spatial resolution of about 150 nm, the LLs will shift gradually in space following the variations in the BS without showing interference at any location. Let us now consider the opposite case of sharp boundaries between domains with different BS. In this situation, at the boundaries, the finite size of our SOT may result in simultaneous detection of LLs originating from the two neighboring domains giving rise to apparent interference. In such a case we expect to observe interference along a network of grain boundaries with width comparable to our SOT size. Instead, Figure 3e shows well defined domains of typical width of 1 µm and length of up to 2 µm, much larger than the SOT size, over which the interference is rather uniform. In addition, most of the domains showing beating are located at the ends or corners of the device, so they do not have two neighboring domains that can cause the apparent interference. Finally, if there is a relative shift in the Dirac point between the neighboring domains, the apparent interference patterns at the boundary would evolve like calculated in Extended Data Fig. 10b, while if $v_F$ changes between the domains the beating node $N_b^1$ of the apparent interference would be independent of $B_a$ as calculated in Extended Data Figs. 10f-i. Both these possibilities are inconsistent with the experimental data. More generally, the $B_a$ dependence of the LL interference due to variations in BS is distinctly different from the one caused by $B_S$. We therefore conclude that disorder that causes spatial variations in BS without creating PMFs cannot explain the observed LL interference.

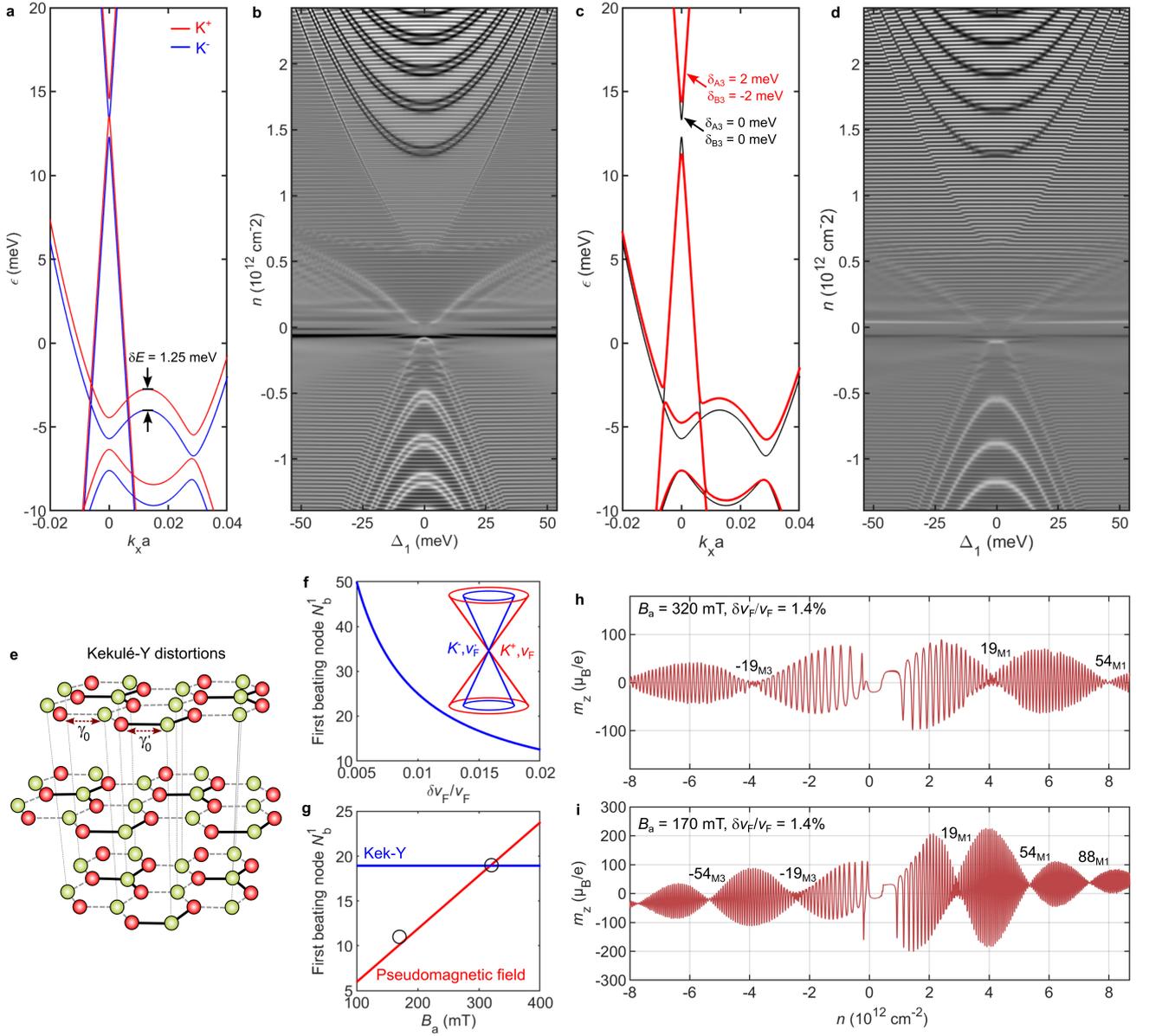

**Extended Data Fig. 10 | Alternative mechanisms of LL interference**. **a,** Band structure of ABA graphene with a relative energy shift of $\delta E = 1.25$ meV between the $K^+$ and $K^-$ bands. **b,** The corresponding calculated QOs show splitting of the lowest MLG LLs inconsistent with the experimental data. **c,** Band structure of ABA graphene with (red) and without (black) staggered sublattice potential $\delta_{A3} = 2$ meV, $\delta_{B3} = -2$ meV. **d,** Corresponding calculated QOs showing no LL interference. **e,** Kekulé-Y distortion with $\gamma_0'$ hopping parameter along the bonds emphasized in black. **f,** The dependence of the first beating node $N_b^1$ on the difference of the Fermi velocities $\Delta v_F / v_F$ of the two valleys. Inset: schematic of the MLG band dispersions of the two valleys. **g,** The dependence of $N_b^1$ on $B_a$ for Kekulé-Y distortion (blue), PMF (red), and the experimental data (circles). **h,** The calculated QOs with $\Delta v_F / v_F = 1.4\%$ at $B_a = 320$ mT. **i,** Same as (h) at $B_a = 170$ mT.

**Acknowledgments**

The authors thank Matan Bocarsly, Yiwen Liu and Bo Han for fruitful discussions. This work was co-funded by the Minerva Foundation grant No 140687, by the National Research Foundation (NRF) of Singapore and Israel Science Foundation (ISF) under ISF-NRF joint program (grants No 3518/20), and by the European Union (ERC, MoireMultiProbe - 101089714). Views and opinions expressed are however those of the author(s) only and do not necessarily reflect those of the European Union or the European Research Council. Neither the European Union nor the granting authority can be held responsible for them. E.Z. acknowledges the support of the Andre Deloro Prize for Scientific Research and Leona M. and Harry B. Helmsley Charitable Trust grant #2112-04911.

**Author contributions**

H.Z. and N.A. developed the setup and performed the scanning measurements. M.U. and W.Z. fabricated the devices and measured the transport. H.Z., Y.Z. and B.Y. conducted the tight-binding calculation and parameter fitting. N.B. and Y.M. fabricated the SOTs and tuning fork assembly. M.E.H. designed and built the SOT readout system. H.Z., N.A. and M.U. performed the data analysis. H.Z., N.A., M.U. and E.Z. wrote the manuscript with contributions from the rest of the authors. K.W. and T.T. provided the hBN crystals.

**Competing interests**

The authors declare no competing interests.

**Data availability**

The data that support the findings of this study are available from the corresponding authors on reasonable request.

**Code availability**

The band structure calculations codes used in this study are available from the corresponding authors on reasonable request.